\UseRawInputEncoding

\documentclass[journal]{IEEEtran}  

\usepackage{multirow}
\usepackage[table,xcdraw]{xcolor}

\usepackage{amsmath,amsfonts}
\usepackage{algorithmic}
\usepackage{array}
\usepackage[caption=false,font=footnotesize,labelfont=rm,textfont=rm]{subfig}
\usepackage{textcomp}
\usepackage{stfloats}
\usepackage{url}
\usepackage{verbatim}
\usepackage{graphicx}

\usepackage{color}
\usepackage{amssymb}
\usepackage{booktabs}
\usepackage[table]{xcolor}
\usepackage{graphicx}
\usepackage{multirow}
\usepackage{color} 
\usepackage{cite}
\usepackage{mathrsfs}

\newcommand{\figref}[1]{Fig. \ref{#1}}

\newcommand{\ra}[1]{\renewcommand{\arraystretch}{#1}}
\newcommand{\subfigref}[2]{Fig. \ref{#1} \subref{#2}}
\newcommand{\tabref}[1]{Table \ref{#1}}

\usepackage{algorithm}
\usepackage{algorithmic}

\newcommand{\alref}[1]{Algorithm \ref{#1}}

\def\BibTeX{{\rm B\kern-.05em{\sc i\kern-.025em b}\kern-.08em
		T\kern-.1667em\lower.7ex\hbox{E}\kern-.125emX}}
\usepackage{balance}
\usepackage{hyperref}
\hypersetup{hidelinks} 
\hypersetup{
	    colorlinks=true,
		linkcolor=blue
	}

\begin{document}
	\title{Channel Fingerprint Construction for Massive MIMO: A Deep Conditional Generative Approach}

\author{
	Zhenzhou~Jin,~\IEEEmembership{Graduate Student Member,~IEEE,}
	Li~You,~\IEEEmembership{Senior Member,~IEEE,}
	Xudong~Li,
	Zhen~Gao,~\IEEEmembership{Senior Member,~IEEE,}
	Yuanwei~Liu,~\IEEEmembership{Fellow,~IEEE,}
	Xiang-Gen~Xia,~\IEEEmembership{Fellow,~IEEE,}
	and~Xiqi~Gao,~\IEEEmembership{Fellow,~IEEE}
	\thanks{Part of this work has been accepted for presentation at the IEEE INFOCOM 2025 \cite{jininfocom}.}	
	\thanks{
		Zhenzhou Jin, Li You, Xudong Li, and Xiqi Gao are with the National Mobile Communications Research Laboratory, Southeast University, Nanjing 210096, China, and also with the Purple Mountain Laboratories, Nanjing 211100, China (e-mail: zzjin@seu.edu.cn; lyou@seu.edu.cn; xdli@seu.edu.cn; xqgao@seu.edu.cn).
		
		Zhen Gao is with the State Key Laboratory of CNS/ATM, Beijing Institute
		of Technology, Beijing 100081, China, also with the Beijing Institute
		of Technology, Zhuhai 519088, China, also with the MIT Key Laboratory
		of Complex-Field Intelligent Sensing, Beijing Institute of Technology, Beijing
		100081, China, also with the Advanced Technology Research Institute, Beijing
		Institute of Technology Jinan, Jinan 250307, China, and also with the Yangtze
		Delta Region Academy, Beijing Institute of Technology Jiaxing, Jiaxing 314019,
		China (e-mail: gaozhen16@bit.edu.cn).
		
		Yuanwei Liu is with the Department of Electrical and Electronic Engineering, The University of Hong Kong, Hong Kong (e-mail: yuanwei@hku.hk).
		
		Xiang-Gen Xia is with the Department of Electrical and Computer Engineering, University of Delaware, Newark, DE 19716, USA (e-mail: xxia@ee.udel.edu).
	}



			
	}

\maketitle

\begin{abstract}
Accurate channel state information (CSI) acquisition for massive multiple-input multiple-output (MIMO) systems is essential for future mobile communication networks. Channel fingerprint (CF), also referred to as channel knowledge map, is a key enabler for intelligent environment-aware communication and can facilitate CSI acquisition. However, due to the cost limitations of practical sensing nodes and test vehicles, the resulting CF is typically coarse-grained, making it insufficient for wireless transceiver design. In this work, we introduce the concept of CF twins and design a \textit{conditional} generative diffusion model (CGDM) with strong implicit prior learning capabilities as the computational core of the CF twin to establish the connection between coarse- and fine-grained CFs. Specifically, we employ a variational inference technique to derive the evidence lower bound (ELBO) for the log-marginal distribution of the observed fine-grained CF \textit{conditioned} on the coarse-grained CF, enabling the CGDM to learn the complicated distribution of the target data. During the denoising neural network optimization, the coarse-grained CF is introduced as \textit{side information} to accurately guide the conditioned generation of the CGDM. To make the proposed CGDM lightweight, we further leverage the additivity of network layers and introduce a one-shot pruning approach along with a multi-objective knowledge distillation technique. Experimental results show that the proposed approach exhibits significant improvement in reconstruction performance compared to the baselines. Additionally, zero-shot testing on reconstruction tasks with different magnification factors further demonstrates the scalability and generalization ability of the proposed approach.



\end{abstract}

\section{Introduction}\label{sec:net_intro}
\IEEEPARstart{T}{he} deep integration of wireless communications, artificial intelligence (AI), and environmental sensing is expected to enable the 6th generation (6G) mobile communication networks to ``perceive'' the physical world with capabilities surpassing human sensing, thereby facilitating the creation of digital twins (DTs) in the virtual realm \cite{9737357,10054381}. The vision of 6G is to propel society towards ``intelligent internet of everything'' and ``ubiquitous connectivity'', realizing the seamless integration and interaction between the physical and virtual worlds. To achieve this, 6G will need to possess more powerful end-to-end information processing capabilities to support emerging applications and domains, including autonomous vehicles, indoor localization, and the metaverse. Therefore, to achieve ultra-low latency and superior performance, while supporting scenarios that integrate AI, sensing, and communication, DT and environmental sensing are considered key enablers for the upcoming 6G era \cite{khan2022digital}.

With the dramatic increase in the antenna array dimensions in massive multiple-input multiple-output orthogonal frequency division multiplexing (MIMO-OFDM) communication systems, the rapid rise in the number and density of user devices, and the utilization of broader bandwidths, 6G networks will encounter the challenge of processing ultra-large-dimensional MIMO channels \cite{10054381,jin2024gdm4mmimo,10403776,jin2024i2i}. Traditional pilot-based methods for acquiring and feedback of channel state information (CSI) may suffer from prohibitively high pilot signaling overhead. Furthermore, traditional wireless transceiver designs typically rely on channel modeling, which is based on specific assumptions and probability distributions of channel parameters. However, these stringent assumptions may not be feasible in high-dynamic, complicated wireless propagation environments \cite{zeng2024tutorial}. As a key determinant, the wireless propagation environment significantly affects channel parameter variation and the performance of wireless communication systems. Consequently, there has been growing interest in environment-aware wireless communications from both the academic and industrial communities.

Channel fingerprint (CF), also referred to as channel knowledge map (CKM), is an emerging enabling technology for environment-aware communications, offering location-specific channel knowledge related to a potential base station (BS) for any BS-to-everything (B2X) pairs \cite{zeng2024tutorial,jin2024i2i}. Ideally, a fine-grained CF serves as a location-specific knowledge base, covering all precise locations within the target communication area. It includes the exact positions of transmitters and receivers, along with their corresponding channel knowledge. This database stores channel-related knowledge for specific locations, including channel power, angle of arrival/departure, and channel impulse responses, which can alleviate the challenges of CSI acquisition and empower the design of wireless transmission technologies. By providing essential and accurate channel knowledge, CF has recently spurred extensive research for various applications in space-air-ground integrated networks, including beam alignment \cite{wu2023environment,zeng2024tutorial}, communication among non-cooperative nodes \cite{zeng2024tutorial}, physical environment sensing \cite{zhang2020constructing, 9838964} and user localization \cite{yang2013rssi}, UAV communication \cite{10173754} and path optimization \cite{9814544,9269485}, resource allocation in air-ground integrated mobility \cite{10556774}.

\begin{figure*}[!t]
	\centering
	\includegraphics[width=0.96\textwidth]{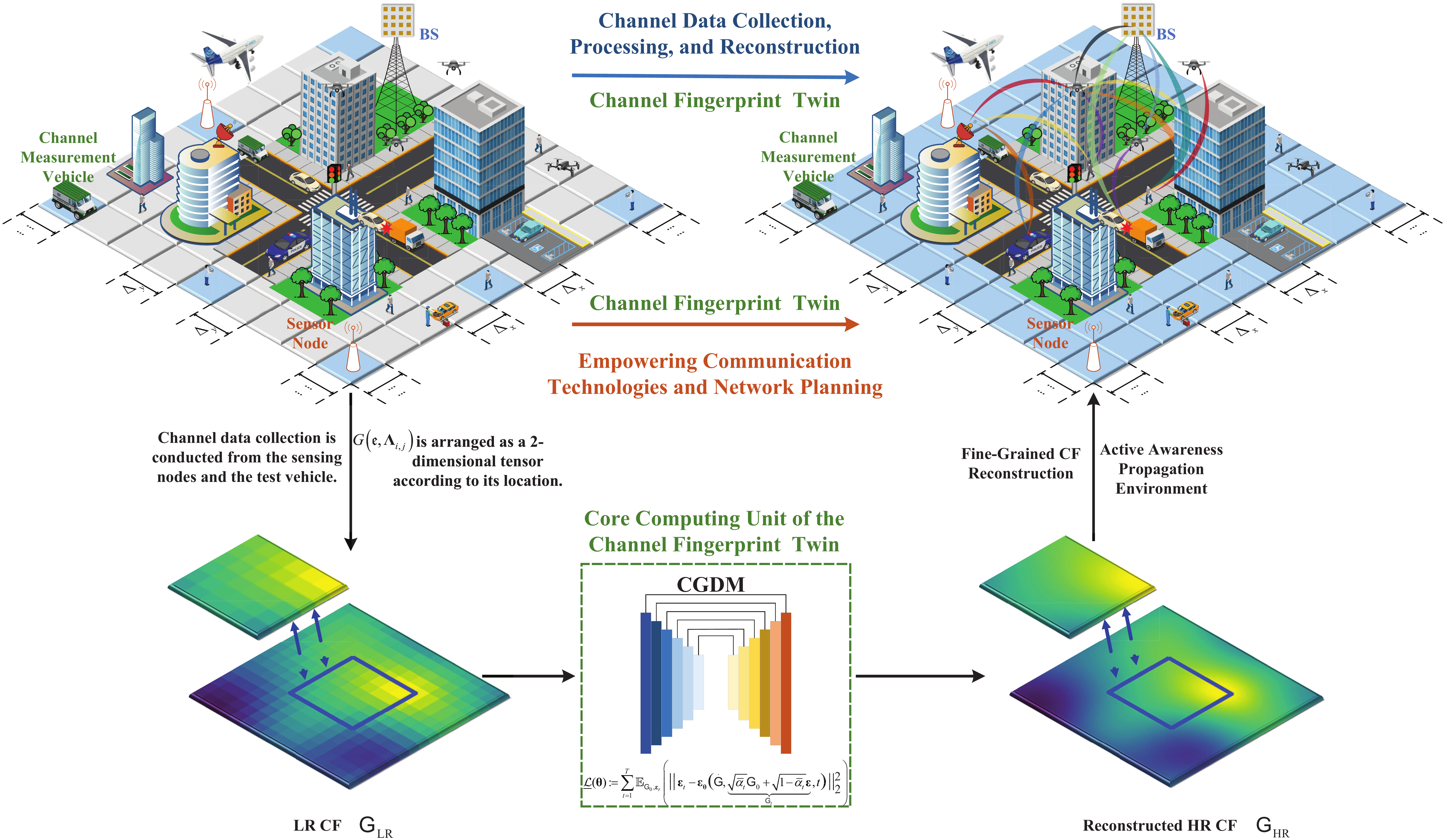}
	\captionsetup{font=footnotesize}
	\caption{Schematic diagram of the CF twin: CGDM functions as the core computational unit of the CF twin, reconstructing fine-grained CF to optimize wireless transmission technologies and network planning.}
	\label{fig:city}
\end{figure*} 

The aforementioned promising applications hinge on the effective construction of CF, which serves as the cornerstone of environment-aware wireless communications. Existing related works can primarily be categorized into model-driven and data-driven approaches. For the former, the authors of \cite{10530520} employ an analytical channel model to represent the spatial variation of the propagation environment, where channel modeling parameters are estimated from measured data to reconstruct the CF. The authors of \cite{8662745} combine prior assumptions of the wireless propagation environment model with partially observed channel measurements to infer channel knowledge at unmeasured locations. For the latter, the work \cite{jin2024i2i} transforms the CF construction task into an image-to-image inpainting problem and designs a Laplacian pyramid-based model to learn the differences between frequency components, enabling efficient reconstruction of the CF for a target area. In \cite{9354041, 9771088}, U-Net is employed to learn geometry-based and physics-based features in urban or indoor environments, thereby constructing the corresponding CFs. In \cite{zeng2024tutorial}, fully connected networks are employed to predict channel knowledge at potential locations using simple 2D coordinate information. Most existing related works focus on constructing CFs using physical propagation environment features or prior assumptions of physical propagation models.

Evidently, a finer-grained CF can assist the BS in acquiring more precise, location-specific channel knowledge \cite{zeng2024tutorial}. However, two key challenges must be addressed for CF to enable environment-aware wireless communications. First, the operation and maintenance costs associated with sensing nodes and test vehicles for measuring channel-related knowledge are inherently constrained. Second, storing a large amount of fine-grained CFs may incur unnecessary and prohibitively high storage overhead for the BS. Consequently, most available CFs are coarse-grained, lacking precise location-specific channel knowledge. Motivated by these practical challenges, we seek to develop a computational unit dedicated to enhancing CF granularity, particularly by reconstructing ultra-fine-grained CFs from coarse-grained counterparts.


%

In this context, fully leveraging measurable channel knowledge and location information from the physical world becomes essential. The concept of DT has emerged as a promising paradigm, widely recognized as a virtual modeling framework that digitally replicates and extends the physical world \cite{khan2022digital}. As illustrated in \figref{fig:city}, from the DT perspective, a structured coarse-grained CF can be sensed and reorganized by sensing nodes and test vehicles deployed in the physical world. This coarse-grained CF represents the physical object, which is subsequently processed by the DT \textit{hub}, serving as the core computing unit of the DT, to generate a fine-grained CF, commonly referred to as the virtual twin object. Leveraging the generated virtual twin facilitates optimized decision-making for subsequent wireless transceiver designs, such as precoding and resource scheduling \cite{zeng2024tutorial, jin2024i2i}. Accordingly, our goal is to establish the intrinsic mapping between physical objects and their corresponding virtual twins. Consistent with the DT paradigm, we refer to this concept as the \textit{CF twins}. To better explore the relationship between the two, we reformulate the task of fine-grained CF construction as an image super-resolution (ISR) problem. However, the conditional distribution of SR outputs given coarse-grained inputs typically follows a complicated parametric distribution, potentially resulting in suboptimal performance for most feedforward neural network-based regression algorithms in ISR tasks \cite{ledig2017photo,Rombach_2022_CVPR}.

Recently, generative AI (GenAI) has emerged as a promising technique for high-fidelity DT construction, showcasing exceptional capabilities in modeling complicated relationships and distributions to effectively synthesize and reconstruct high-dimensional data \cite{Ho11,ledig2017photo,Rombach_2022_CVPR,JINTCCN,du2024enhancing,kingma2013auto}. Among various GenAI techniques, generative diffusion models (GDM) \cite{Ho11} are widely recognized as one of the most prominent classes of generative models. GDM does not require training additional discriminators, as in generative adversarial network (GAN) \cite{ledig2017photo}, aligning posterior distributions like variational autoencoders (VAE) \cite{kingma2013auto}. Therefore, GDMs have shown remarkable performance in high-dimensional data generation tasks \cite{lovelace2024latent,Rombach_2022_CVPR}, underscoring their versatile application potential and efficacy. However, conventional GDMs are typically unconditional, which may lead to uncertainty in accurately generating the target fine-grained CF, potentially resulting in multiple possible solutions for the target CF.


To this end, this paper proposes a \textit{conditional} GDM (CGDM) with powerful implicit prior learning capabilities, serving as the CF twin hub, where coarse-grained CF is introduced as side information to guide the iterative refinement process of the denoising neural network. Moreover, to make the proposed CGDM lightweight, we further introduce an efficient pruning approach and a multi-objective knowledge distillation technique. Specifically, the main contributions of this paper are as follows:
\begin{itemize}
	\item Building on the proposed concept of CF twin, we treat the coarse-grained CF as the physical object and the fine-grained CF as its corresponding virtual twin. To effectively capture the relationship between them, we reformulate this task as an ISR problem and solve it utilizing a Stein score-based iterative refinement approach.

	\item To enable the CF twin hub to more effectively learn the connections between physical entities and their virtual counterparts, we adopt a variational inference technique to derive the evidence lower bound (ELBO) of the log-\textit{conditional} marginal distribution of the observed fine-grained CF, which serves as a surrogate training objective. Maximizing this objective allows CGDM to learn the true target CF distribution, facilitating the transition from a standard normal distribution to the desired distribution. To better guide this transition, the coarse-grained CF is incorporated as side information during the optimization of the denoising network, thereby enhancing the controllability and fidelity of fine-grained CF generation.

	\item 
	Considering that the proposed CGDM is a large AI model with numerous parameters, we develop an efficient pruning approach to enable its lightweight deployment. Specifically, we formulate the layer pruning task as a combinatorial optimization problem. By leveraging the inherent additivity property of network layers, we introduce a one-shot layer pruning strategy along with a multi-objective knowledge distillation technique, resulting in the lightweight CGDM (LiCGDM).

	\item 
	Experimental results show that the proposed approach achieves significant performance improvements over the baselines. Additionally, we validate the generalization and knowledge transfer capabilities of the proposed model by conducting zero-shot performance testing on other SR CF tasks with unseen magnification factors (e.g., $\times16$, $\times8$).

	
	

\end{itemize}

The rest of this paper is structured as follows. Section II introduces the system model and the formulation of the fine-grained CF construction problem. The overall design of the CGDM and its specific network architecture are introduced in Section III. The one-shot pruning approach and knowledge distillation technique are proposed in Section IV. Experimental results are presented in Section V, with the conclusions provided in Section VI.

\section{System Model And Problem Formulation}\label{sec:sys_mod}
In this section, we first outline the communication scenario and present the massive MIMO-OFDM physical channel model for each potential user equipment (UE) location. Then, we model the specific CF, that is, the channel power, for each potential UE based on its location and environmental factors. Finally, we describe the fine-grained CF reconstruction problem from the perspective of ISR.

\subsection{Massive MIMO Physical Channel Model}
Consider a massive MIMO-OFDM communication scenario within a square area $A\subset {\mathbb{R}^2}$, where the 2D coordinates of the UE locations are defined as $\left\{ {{{{\mathbf{x}}_m}}} \right\}_{m = 1}^M = {\mathcal{A}}$. Specifically, the BS is outfitted with a uniform planar array (UPA) consisting of $N_r=N_{r,v}\times N_{r,h}$ antenna elements, spaced at half-wavelength intervals, and serves $M$ single-antenna users within a cell. Here, $N_{r,v}$ and $N_{r,h}$ represent the numbers of antennas in each vertical column and horizontal row, respectively. It is assumed that the wireless propagation environment contains ${L_{{{\mathbf{x}}_m}}}$ physical paths from the BS to the UE located at ${{\mathbf{x}}_m}$. The system employs OFDM modulation with $N_c$ subcarriers, characterized by an adjacent subcarrier spacing of $\Delta_{f}$. Typically, $N_k$ active subcarriers are utilized at the center of the total $N_c$ subcarriers for signal transmission, while the remaining subcarriers serve as guard bands. Define the set of active subcarriers as $\mathcal{K}\!=\!\left\lbrace -\frac{N_{k}}{2},-\frac{N_{k}}{2}+1,\ldots,\frac{N_{k}}{2}-1\right\rbrace $. The spatial-frequency domain response of the channel between the BS and the UE over active subcarriers $\mathcal{K}$ is modeled as \cite{liu2023structured}
\begin{align}\label{eq:1}
	{\mathbf{h}}^{\rm{H}}\!\left({{\mathbf{x}}_m} \right)\! \!=\!\!\!\sum_{l=1}^{L_{{{\mathbf{x}}_m}}}\!\!\alpha_{l,{{{\mathbf{x}}_m}}}\mathbf{a}_t^{\rm{H}}\!\left(\tau_{l,{{{\mathbf{x}}_m}}}\right)\!\otimes\!\mathbf{a}_h^{\rm{H}}\left(\psi_{l,{{{\mathbf{x}}_m}}}\right)\!\otimes\!\mathbf{a}_v^{\rm{H}}\left(\phi_{l,{{{\mathbf{x}}_m}}}\right),
\end{align}
where $\otimes$ denotes the Kronrcker product, and $\alpha_{l,{{{\mathbf{x}}_m}}}$ and $\mathbf{a}_t\left(\tau_{l,{{{\mathbf{x}}_m}}}\right)$ represent the complex gain and the propagation delay, respectively, corresponding to the $l$-th path for the UE located at ${{{{\mathbf{x}}_m}}}$. The normalized horizontal angle \(\psi_{l,{{{\mathbf{x}}_m}}} \in [-1, 1]\) is related to the elevation angle \(\theta_{l,{{{\mathbf{x}}_m}}}\), while the normalized vertical angle \(\phi_{l,{{{\mathbf{x}}_m}}} \in [-1, 1]\) is associated with the elevation angle \(\theta_{l,{{{\mathbf{x}}_m}}} \in \left[-\frac{\pi}{2}, \frac{\pi}{2}\right]\) through the relationship \(\phi_{l,{{{\mathbf{x}}_m}}} = \sin \theta_{l,{{{\mathbf{x}}_m}}}\). The azimuth angle \(\vartheta_{l,{{{\mathbf{x}}_m}}} \in \left(-\frac{\pi}{2}, \frac{\pi}{2}\right)\) is defined by \(\psi_{l,{{{\mathbf{x}}_m}}} = \cos \theta_{l,{{{\mathbf{x}}_m}}} \sin \vartheta_{l,{{{\mathbf{x}}_m}}}\). The steering vectors $\mathbf{a}_t\left(\tau_{l,{{{\mathbf{x}}_m}}}\right)$,$\mathbf{a}_h\left(\psi_{l,{{{\mathbf{x}}_m}}}\right)$, and $\mathbf{a}_v\left(\phi_{l,{{{\mathbf{x}}_m}}}\right)$ can be represented as
\begin{align}
	\mathbf{a}_v\left(\phi_{l,{{\mathbf{x}}_m}}\right)=\left[1,e^{-\jmath\pi\phi_{l,{{\mathbf{x}}_m}}},\ldots,e^{-\jmath\pi(N_{r,v}-1)\phi_{l,{{\mathbf{x}}_m}}}\right]^T,
\end{align}
\begin{align}
	\mathbf{a}_h\left(\psi_{l,{{\mathbf{x}}_m}}\right)=\left[1,e^{-\jmath\pi\psi_{l,{{\mathbf{x}}_m}}},\ldots,e^{-\jmath\pi\left(N_{r,h}-1\right)\psi_{l,{{\mathbf{x}}_m}}}\right]^T,
\end{align}
\begin{align}
	\mathbf{a}_{t}\left(\tau_{l,{{\mathbf{x}}_m}}\right)\!=\!\left[\!e^{\!-\!\jmath2\pi\left(\!-\!\frac{N_{k}}{2}\!\right)\!\Delta_{f}\tau_{l,{{\mathbf{x}}_m}}}\!,\!\cdots\! ,e^{\!-\!\jmath2\pi\left(\!\frac{N_{k}}{2}\!-\!1\!\right)\!\Delta_{f}\tau_{l,{{\mathbf{x}}_m}}}\!\right]^{T}\!\!\!,\!
\end{align}
where $\mathbf{a}_{h}(\psi_{l,{{\mathbf{x}}_m}})\in\mathbb{C}^{N_{r,h}\times1}$, $	\mathbf{a}_v\left(\phi_{l,{{\mathbf{x}}_m}}\right)\in\mathbb{C}^{N_{v,h}\times1}$, and $\mathbf{a}_{t}\left(\tau_{l,{{\mathbf{x}}_m}}\right)\in\mathbb{C}^{N_{k}\times1}$. 

\subsection{Channel Fingerprints Model}
Based on the proposed physical channel model in Section II.A, the channel power at the UE located at $\mathbf{x}_m$, in dB scale, is defined as 
\begin{align}\label{eq:}
	{G}\left(\mathfrak{e},\mathbf{x}_m\right) = {\left( {\mathbb{E}\left\{ { {{{\mathbf{h}}^{\text{H}}}\left( {{{\mathbf{x}}_m}} \right){\mathbf{h}}\left( {{{\mathbf{x}}_m}} \right)} } \right\} } \right)_{{\rm{dB}}}},
\end{align} 
where $\mathbb{E}\left\{  \cdot  \right\}$ denotes the expectation operation, and $\mathfrak{e}$ represents the propagation environment, which determines the channel characteristics in \eqref{eq:1}, including the complex gain and propagation delay. It is evident that the channel power attenuation experienced at the UE is influenced by various environmental factors, including propagation losses along different paths as well as reflections and diffractions caused by surrounding structures such as buildings \cite{9354041,jin}. In this paper, we refer to the collection of channel power values at potential locations as the unstructured CF.

Since the communication area under consideration is a 2D square area $A\subset {\mathbb{R}^2}$, we can perform spatial discretization along both the $X$-axis and $Y$-axis. Specifically, given an area of interest with size $\mathcal{W}\times\mathcal{W} $, we define a resolution factor $\sigma$, such that the minimum spacing units in the spatial discretization process are $\Delta_x={\mathcal{W} \mathord{\left/
		{\vphantom {\mathcal{W} \sigma}} \right.
		\kern-\nulldelimiterspace} \sigma}$ and $\Delta_y={\mathcal{W} \mathord{\left/
		{\vphantom {\mathcal{W} \sigma}} \right.
		\kern-\nulldelimiterspace} \sigma}$. Each spatial grid is located at ${{\mathbf{\Lambda }}_{i,j}}$, where $i = 1,2,...,	{\mathcal{W} \mathord{\left/
		{\vphantom {\mathcal{W} \Delta_x}} \right.
		\kern-\nulldelimiterspace} \Delta_x}$ and $j = 1,2,...,	{\mathcal{W} \mathord{\left/
		{\vphantom {\mathcal{W} \Delta_y}} \right.
		\kern-\nulldelimiterspace} \Delta_y}$, and the $\left( i, j\right)$-th spatial grid can be represented as 
\begin{align}\label{eq:}
	{{\mathbf{\Lambda }} _{i,j}}: = {[i{\Delta _x},j{\Delta {}_y]^T}}.
\end{align}
Given the resolution factor $\sigma$, the unstructured CF corresponding to potential UE locations within the target area can be rearranged into a two-dimensional tensor, denoted as 
\begin{align}
 {[\boldsymbol{\mathsf{G}}]_{i,j}} = { G}\left(\mathfrak{e}, {{\mathbf{\Lambda }}_{i,j}}\right).
\end{align}
Furthermore, by incorporating additional dimensions such as time and frequency, the CF model can be naturally extended to a higher-order tensor representation.

\textit{Remark 1}: When the target area has a size of $\mathcal{W}\times\mathcal{W}$ and the resolution factor is $\sigma$, the number of points requiring interpolation is $\lceil{\mathcal{W}}/{\sigma}\rceil^2$. As the resolution factor $\sigma$ increases, the complexity of traditional interpolation algorithms also grows with ${\cal O}( \lceil{ {\mathcal{W}}/{\sigma}\rceil}^3) $, posing significant challenges for constructing fine-grained CF in practical scenarios. Alternatively, the spatial resolution of the CF can be improved by increasing the density of measurement points, either through deploying more sensing nodes or by employing test vehicles to collect channel knowledge at finer geographical intervals. However, both approaches may be impractical due to high hardware and labor costs.

\subsection{Problem Formulation}
Define a coarse-grained factor ${\sigma}_{\rm{LR}}$ and a fine-grained factor ${\sigma}_{\rm{HR}}$, where ${\sigma}_{\rm{HR}}$ is typically an integer multiple (e.g., $\times4$, $\times8$) of ${\sigma}_{\rm{LR}}$. Accordingly, the coarse-grained CF and fine-grained CF are denoted by ${\boldsymbol{\mathsf{G}}}_{\rm{LR}}$ and ${\boldsymbol{\mathsf{G}}}_{\rm{HR}}$, which are collected by discretizing the target area into $\sigma_{\mathrm{LR}}\times \sigma_{\mathrm{LR}}$ and $\sigma_{\mathrm{HR}}\times \sigma_{\mathrm{HR}}$ grids, respectively. In light of \textit{Remark 1}, the \textit{CF twin} aims to reconstruct the fine-grained CF ${\boldsymbol{\mathsf{G}}}_{\rm{HR}}$ from a given coarse-grained CF ${\boldsymbol{\mathsf{G}}}_{\rm{LR}}$, particularly in scenarios constrained by measurement costs, privacy concerns, or security requirements.

In a typical ISR task, the goal is to reconstruct a high-resolution (HR) image from a given low-resolution (LR) counterpart, thereby enhancing the fine details and overall quality of the image. It can be observed that our problem aligns with the ISR task, which inspires us to analyze the fine-grained CF reconstruction problem from an ISR perspective. Specifically, we treat the elements of the $\boldsymbol{\mathsf{G}}_{\rm{LR}}$ matrix as pixels, viewing the coarse-grained $\boldsymbol{\mathsf{G}}_{\rm{LR}}$ as an LR image and the fine-grained $\boldsymbol{\mathsf{G}}_{\rm{HR}}$ as an HR image. Then, our goal is to learn a specific mapping relationship that efficiently reconstructs HR CF $\boldsymbol{\mathsf{G}}_{\rm{HR}}$ from a given LR CF $\boldsymbol{\mathsf{G}}_{\rm{LR}}$, i.e.,
\begin{align}\label{eq:11}
	{\mathcal{M}_\Theta }:\boldsymbol{\mathsf{G}}_{{\rm{LR}},u} \to \boldsymbol{\mathsf{G}}_{{\rm{HR}},u},\quad \forall u \in \left\{ {1,2,\ldots,U} \right\},
\end{align}
where $\Theta$ is the parameter set of this mapping $\mathcal{M}_\Theta$, and $U$ is the number of training samples. However, this task represents a classic and highly challenging inverse problem, requiring the effective reconstruction of fine details from a given LR CF. Since the conditional distribution of HR outputs corresponding to a given LR input typically does not adhere to a simple parametric distribution, many feedforward neural network-based regression methods for ISR tend to perform poorly at higher upscaling factors, struggling to recover fine details accurately \cite{Rombach_2022_CVPR}. In contrast, deep generative models have demonstrated success in learning complicated empirical distributions of target data. Specifically, if the implicit prior information of the HR CF distribution, such as the gradient of the data log-density, can be learned, one can transition to the target CF distribution through iterative sampling steps from a standard normal distribution, similar to Langevin dynamics \cite{song2019generative}. Therefore, learning the target HR CF distribution can be solved by optimizing
\begin{subequations}\label{eq:1212}
	\begin{align}
		&\!\!\!\!\mathop {{\text{argmin}}}\limits_\Theta  {{\mathbb{E}}_{{p}(\boldsymbol{\mathsf{G}}_\mathrm{HR})}}\!\left[ {{{\!\left\| {\nabla \log {p}(\boldsymbol{\mathsf{G}}_\mathrm{HR}) \!-\! \nabla \log {p_{\Theta}}(\boldsymbol{\mathsf{G}}_\mathrm{HR})} \right\|}^2_2}} \right] ,\label{eq:dbso}\\
		&\!\!\!\!\qquad\qquad\quad{\rm{s.t.}}\text{  } {\mathbf{x}_m} \in {\mathcal{A}} ,u \in \{1,2,\ldots, U\},\label{eq:}
	\end{align}
\end{subequations}
where $\nabla \log {p}(\boldsymbol{\mathsf{G}}_\mathrm{HR})$ represents the gradient of the HR CF log-density, also referred to as the Stein score, and $p_{\Theta}$ denotes the learned density. It has been shown that the noise learned by traditional GDM is equivalent to the Stein score \cite{Ho11}, enabling the generation of HR CF samples aligned with the target data distribution. However, for traditional GDMs, accurately generating the target HR CF is an ill-posed task, meaning that the generation process may yield multiple possible solutions for the HR CF. Therefore, in contrast to traditional GDMs, which starts with a pure Gaussian noise tensor, introducing an additional source signal as \textit{side information} (also guiding condition) is essential to achieve an optimal solution. In this context, the objective \eqref{eq:1212} needs to be reformulated as
\begin{subequations}\label{eq:13}
	\begin{align}
		&\!\!\!\mathop {{\text{argmin}}}\limits_\Theta  {{\mathbb{E}}_{{p}(\!\boldsymbol{\mathsf{G}}_\mathrm{H\!R},\boldsymbol{\mathsf{G}}_\mathrm{L\!R}\!)\!\!}}\!\left[ {{{\!\left\| {\nabla \!\log {p}(\!\boldsymbol{\mathsf{G}}_\mathrm{H\!R}|\boldsymbol{\mathsf{G}}_\mathrm{L\!R}\!) \!\!-\!\!\! \nabla \!\log {p_{\Theta}}(\!\boldsymbol{\mathsf{G}}_\mathrm{H\!R}|\boldsymbol{\mathsf{G}}_\mathrm{L\!R}\!)} \right\|}^2_2}} \right]\!\!,\!\label{eq:dbso}\\
		&\!\!\!\qquad\qquad\quad{\rm{s.t.}}\text{  } {\mathbf{x}_m} \in {\mathcal{A}} ,u \in \{1,2,\ldots, U\}.\label{eq:}
	\end{align}
\end{subequations}
To this end, our approach leverages the prior information learned, with the LR CF $\boldsymbol{\mathsf{G}}_{\rm{LR}}$ serving as the additional source signal to guide the iterative refinement process. Further implementation details are provided in Section \ref{sec:CGDM}.



\section{CGDM-Enabled SR CF}\label{sec:CGDM}
In this section, we introduce the proposed CGDM as the core computational unit of the CF twin. First, under the variational inference framework, we derive a concrete proxy objective to ensure the effective operation of the proposed CGDM. Next, we introduce the LR CF as side information and design a \textit{conditional} GDM to iteratively refine the transformation from a standard normal distribution to the target data distribution, akin to Langevin dynamics, for HR CF reconstruction. Finally, we present the detailed network architecture of the proposed CGDM. To simplify the notation, ${\boldsymbol{\mathsf{G}}_{{\rm{LR}}}}$ and ${\boldsymbol{\mathsf{G}}_{{\rm{HR}}}}$ are represented by ${ \boldsymbol{\dot{\mathsf{G}}}}$ and $\boldsymbol{\mathsf{G}}$, respectively, in the subsequent sections.

\subsection{CGDM Design for SR CF}
Given a dataset of LR CF inputs paired with HR CF outputs, defined as $\mathcal{D}\!=\!\lbrace {{ \boldsymbol{\dot{\mathsf{G}}}}_u,\boldsymbol{\mathsf{G}}_u}\rbrace_{u=1}^U $, which represent samples drawn from an unknown distribution $p({ \boldsymbol{\dot{\mathsf{G}}}},\boldsymbol{\mathsf{G}} )$. Such datasets are generally collected from sensing nodes and test vehicles, with different resolution factors (e.g., ${\sigma}_{\rm{LR}}$ and ${\sigma}_{\rm{HR}}$), depending on the specific scenario. In our task, the focus is on learning a parametric approximation to $p( {\boldsymbol{\mathsf{G}}}| { \boldsymbol{\dot{\mathsf{G}}}} ) $ through a directed iterative refinement process guided by source information, which enables the mapping of ${ \boldsymbol{\dot{\mathsf{G}}}}$ to ${\boldsymbol{\mathsf{G}}}$. Given the powerful implicit prior learning capability of GDM, we design a \textit{conditional} GDM to facilitate the generation of $\boldsymbol{\mathsf{G}}$.

\begin{figure}[!t]
	\centering
	\includegraphics[width=0.485\textwidth]{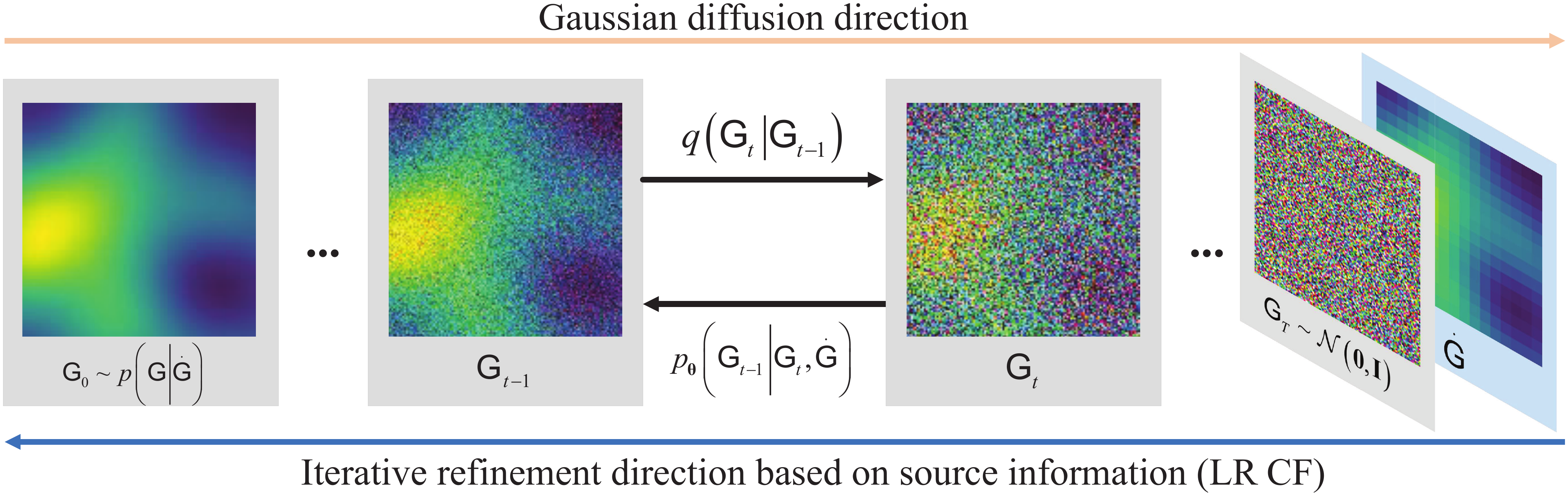}
	\captionsetup{font=footnotesize}
	\caption{The mechanism of CGDM for generating HR CF consists of a Gaussian diffusion process (without learnable parameters) and an iterative refinement process based on LR CF. Specifically, the pink arrow indicates the direction of the Gaussian diffusion process, which progressively adds noise to the HR CF. The blue arrow indicates the direction of the iterative refinement process, which utilizes the implicit prior learned during training and is conditioned on the source information (LR CF) to generate the HR CF.}
	\label{fig:generative_process}
\end{figure}

As shown in \figref{fig:generative_process}, CGDM can generate the target HR CF defined as ${\boldsymbol{\mathsf{G}}}_0$ through $T$ refinement time steps. Beginning with a CF ${\boldsymbol{\mathsf{G}}}_T \sim \mathcal{N}\left( {\mathbf{0},\mathbf{I}} \right)$ composed of pure Gaussian noise, CGDM iteratively refines this initial input based on the source signal and the prior information learned during the training process, namely the \textit{conditional} distribution ${p_{{\boldsymbol{\theta}}} }( {{{\boldsymbol{\mathsf{G}}}_{t - 1}}| {{{\boldsymbol{\mathsf{G}}}_t},{ \boldsymbol{\dot{\mathsf{G}}}}} }) $. As it progresses through each time step $t$, it produces a series of output CFs defined as $\left\lbrace {\boldsymbol{\mathsf{G}}}_{T-1}, {\boldsymbol{\mathsf{G}}}_{T-2},...,{\boldsymbol{\mathsf{G}}}_{0} \right\rbrace $, ultimately resulting in the target HR CF ${\boldsymbol{\mathsf{G}}}_0 \sim p({\boldsymbol{\mathsf{G}}} | { \boldsymbol{\dot{\mathsf{G}}}} ) $. Specifically, the distribution of intermediate CFs in the iterative refinement chain is governed by the forward diffusion process, which gradually adds noise to the output CF through a fixed Markov chain, denoted as ${q( {{\boldsymbol{\mathsf{G}}_t}| {{\boldsymbol{\mathsf{G}}_{t - 1}}} } )}$. Our model seeks to reverse the Gaussian diffusion process by iteratively recovering the signal from noise through a reverse Markov chain conditioned on the source CF ${ \boldsymbol{\dot{\mathsf{G}}}}$. To achieve this, we learn the reverse chain by leveraging a denoising neural network ${{{{{\boldsymbol{\varepsilon}}_{\boldsymbol{\theta }}} }}}\left( \cdot \right)$, optimized utilizing the objective function \eqref{eq:46}. The CGDM takes as input an LR CF and a noisy image to estimate the noise, and after $T$ refinement steps, generates the target HR CF.

\begin{algorithm}[!b]
	\caption{Offline Training Strategy for the Denoising Neural Network \textit{Conditioned} on LR CF}
	\label{alg:train}
	\begin{algorithmic}[1]
		\REPEAT
		\STATE
		Load CF training data pairs $\left({ \boldsymbol{\dot{\mathsf{G}}}},\boldsymbol{\mathsf{G}}_0 \right) \sim p\left({ \boldsymbol{\dot{\mathsf{G}}}},\boldsymbol{\mathsf{G}}_0 \right) $ 
		\STATE
		Obtain time steps $t \sim {\rm{Uniform}}({1,...,T})$
		\STATE
		Randomly generate a noise tensor with the same dimensions as $\boldsymbol{\mathsf{G}}_0$, ${\boldsymbol{\varepsilon}_t} \sim \mathcal{N}\left( {\mathbf{0},\mathbf{I}} \right)$
		\STATE
		Add noise incrementally to the HR CF $\boldsymbol{\mathsf{G}}_0$ according to \eqref{eq:ht} to perform the diffusion process
		\STATE Input the corrupted HR CF ${{{\boldsymbol{\mathsf{G}}}_t}}$, LR CF ${\boldsymbol{\dot{\mathsf{G}}}}$, and time step $t$ into the model ${{{{{\boldsymbol{\varepsilon}}_{\boldsymbol{\theta }}} }}}\left( \cdot \right)$
		\STATE
		Perform gradient descent step on the objective function \eqref{eq:46} to update the model parameters ${\boldsymbol{\theta }}$:\\
		\quad $ \nabla_{\boldsymbol{\theta }} \big\| {{\boldsymbol{\varepsilon }}_t} - {{\boldsymbol{\varepsilon }}_{\boldsymbol{\theta }}}\big({{ \boldsymbol{\dot{\mathsf{G}}}}}, {\sqrt {{{\bar \alpha }_t}} {{{\boldsymbol{\mathsf{G}}}_0}} \!+\! \sqrt {1 - {{\bar \alpha }_t}} {{\boldsymbol{\varepsilon }}}},t \big) \big\|_2^2 $
		\UNTIL the objective function \eqref{eq:46} converges
	\end{algorithmic}
\end{algorithm}

\begin{algorithm}[!t]
	\caption{Inferring the HR CF in the reverse process \textit{conditioned} on the LR CF through $T$ iterative refinement steps}
	\label{alg:infer}
	\begin{algorithmic}[1]
		\STATE
		Load the pretrained model and its weights ${\boldsymbol{\theta }}$
		\STATE
		Obtain the completely corrupted HR CF ${{{\boldsymbol{\mathsf{G}}}_T}}\sim \mathcal{N}\left( {\mathbf{0},\mathbf{I}} \right)$, and LR CF ${\boldsymbol{\dot{\mathsf{G}}}}$
		\FOR{$t=T,..,1$}
		\STATE
		${\boldsymbol{\varepsilon}^*} \sim\mathcal{N}\left( {\mathbf{0},\mathbf{I}} \right)$ if $t>1$, else ${\boldsymbol{\varepsilon}^*} =0$
		\STATE
		Execute the refinement step according to \eqref{eq:48}:\\
		\quad ${{{\boldsymbol{\mathsf{G}}}_{t\!-\!1}}} \!\leftarrow\! \frac{1}{{\sqrt {{\alpha _t}} }}\!\!\left(\!\! {{{{\boldsymbol{\mathsf{G}}}}_t} \!\!-\!\! \frac{{1 - {\alpha _t}}}{{\sqrt {1 - {{\bar \alpha }_t}} }}{{\boldsymbol{\varepsilon}}_{\boldsymbol{\theta }}}\!\left(\!{{ \boldsymbol{\dot{\mathsf{G}}}}, {{\boldsymbol{\mathsf{G}}}_t},t} \!\right)} \!\!\right) \!+\! \sqrt { \frac{{1 \!-\! {{\bar \alpha }_{t \!-\! 1}}}}{{1 \!-\! {{\bar \alpha }_t}}}{\beta _t}} {\boldsymbol{\varepsilon}^*}$
		\ENDFOR
		\RETURN HR CF ${{\hat{{\boldsymbol{\mathsf{G}}}}_0}}$
		
	\end{algorithmic}
\end{algorithm}





\subsection{Diffusion Process Starting with HR CF }
Consider an HR CF sample that is drawn from a distribution of interest, denoted as ${\boldsymbol{\mathsf{G}}_0}={\boldsymbol{\mathsf{G}}} \sim q( \boldsymbol{\mathsf{G}})$. The GDM utilizes a fixed diffusion process $q( {{\boldsymbol{\mathsf{G}}_{1:T}}| {{\boldsymbol{\mathsf{G}}_0}} } )$ for training, which involves relatively high-dimensional latent variables. This process defines a forward diffusion mechanism, a deterministic Markovian chain, where Gaussian noise is gradually introduced to the sample over $T$ steps. The noise level at each step is determined by a variance schedule $\{ {{\beta _t} \in ( {0,1} )} \}_{t = 1}^T$. Specifically, the forward diffusion process is defined as \cite{Ho11}
\begin{align}\label{eq:chmd}
	q\left( {{\boldsymbol{\mathsf{G}}_{1:T}}\left| {{\boldsymbol{\mathsf{G}}_0}} \right.} \right) &= \prod\limits_{t = 1}^T {q\left( {{\boldsymbol{\mathsf{G}}_t}\left| {{\boldsymbol{\mathsf{G}}_{t - 1}}} \right.} \right)}.
\end{align}
Therefore, the forward diffusion process does not involve trainable parameters. Instead, it is a fixed and predefined linear Gaussian model, which can be denoted as
\begin{align}
	\label{eq:marginal ddpm}
	q\left( {{\boldsymbol{\mathsf{G}}_t}\left| {{\boldsymbol{\mathsf{G}}_{t - 1}}} \right.} \right) &= \mathcal{N}\left( {{\boldsymbol{\mathsf{G}}_t};\sqrt {1 - {\beta _t}} {\boldsymbol{\mathsf{G}}_{t - 1}},{\beta _t}\mathbf{I}} \right),  \\
	\label{eq:Gt}
	{\boldsymbol{\mathsf{G}}_t} 
	&= \sqrt {1 - {\beta _t}} {\boldsymbol{\mathsf{G}}_{t - 1}} + \sqrt {{\beta _t}} {\boldsymbol{\varepsilon}},
\end{align}
where $\boldsymbol{\varepsilon}$ denotes Gaussian noise with a distribution of $\mathcal{N}( {\boldsymbol{\varepsilon}; {\mathbf{0},\mathbf{I}} } )$. Define ${\alpha _t} = 1 - {\beta _t}$ and ${\overline \alpha  _t} = \prod\nolimits_{i = 1}^t {{\alpha_i}}$. Under the linear Gaussian assumption of the transition density $q( {{\boldsymbol{\mathsf{G}}_t}| {{\boldsymbol{\mathsf{G}}_{t - 1}}} } )$, and by combining \eqref{eq:marginal ddpm} and \eqref{eq:Gt}, we can utilize the reparameterization technique to sample $\boldsymbol{\mathsf{G}}_t$ in closed form at any given time step $t$:
\begin{align}
	\label{eq:ht}
	{{\boldsymbol{\mathsf{G}}}_t} 
	&=\sqrt {{{\overline \alpha  }_t}} {{\boldsymbol{\mathsf{G}}}_0} + \sqrt {1 - {{\overline \alpha  }_t}} \boldsymbol{\varepsilon}, \\
    \label{eq:chmd}
	q\left( {{\boldsymbol{\mathsf{G}}_t}\left| {{\boldsymbol{\mathsf{G}}_0}} \right.} \right) &= \mathcal{N}\left( {{\boldsymbol{\mathsf{G}}_t};\sqrt {{{\overline \alpha  }_t}} {\boldsymbol{\mathsf{G}}_0},\left(1 - {{\overline \alpha  }_t}\right)  \mathbf{I}} \right).
\end{align}
Generally, the variance schedule is generally set as ${\beta _1} < {\beta _2} < ... < {\beta _T}$. When ${\beta _T}$ is set infinitely close to 1, ${{\boldsymbol{\mathsf{G}}}_T}$ converges to a standard Gaussian distribution for any initial state ${{{\boldsymbol{\mathsf{G}}}_0}}$, i.e., $q( {{{\boldsymbol{\mathsf{G}}}_T}| {{{\boldsymbol{\mathsf{G}}}_0}} } ) \approx \mathcal{N}( {{{\boldsymbol{\mathsf{G}}}_T};\mathbf{0},{\mathbf{I}}} )$.

\subsection{Reverse Diffusion Process Conditioned on LR CF}
For the traditional GDM \cite{Ho11}, the reverse process can be considered a decoding procedure, where at each time step $t$, ${\boldsymbol{\mathsf{G}}_{t}}$ is denoised and restored to ${\boldsymbol{\mathsf{G}}_{t-1}}$, with the transition probability for each step denoted as ${p( {{\boldsymbol{\mathsf{G}}_{t-1}}| {{\boldsymbol{\mathsf{G}}_{t}}} } )}$. Based on the Markov density transition properties, the joint distribution of the reverse process is expressed as 
\begin{align}\label{eq:}
	p\left( {{{{\boldsymbol{\mathsf{G}}}}_{0:T}}} \right){\text{ }} = p\left( {{{{\boldsymbol{\mathsf{G}}}}_T}} \right)\prod\limits_{t = 1}^T {p\left( {{{{\boldsymbol{\mathsf{G}}}}_{t - 1}}\left| {{{{\boldsymbol{\mathsf{G}}}}_t}} \right.} \right)}.
\end{align}
However, deriving the expression for $p( {{{\boldsymbol{\mathsf{G}}}_{t - 1}}| {{{\boldsymbol{\mathsf{G}}}_t}} } )$ utilizing Bayes' theorem reveals that its denominator contains an integral, which lacks a closed-form solution. Consequently, a denoising neural network with parameters ${{\boldsymbol{{\theta}}}}$ needs to be developed to approximate these conditional probabilities in order to execute the reverse process. Since the reverse process also functions as a Markov chain, it can be represented as
\begin{align}
	\label{eq:29}
	{p}\left( {{{\boldsymbol{\mathsf{G}}}_{0:T}}} \right) &= {p}\left( {{{\boldsymbol{\mathsf{G}}}_T}} \right)\prod\limits_{t = 1}^T {{p_{\boldsymbol{\theta }}}\left( {{{\boldsymbol{\mathsf{G}}}_{t - 1}}\left| {{{\boldsymbol{\mathsf{G}}}_t}} \right.} \right)},\\
	\label{eq:30}
	{p_{{\boldsymbol{\theta}}} }\left( {{{\boldsymbol{\mathsf{G}}}_{t - 1}}\left| {{{\boldsymbol{\mathsf{G}}}_t}} \right.} \right) &= \mathcal{N}\left( {{{\boldsymbol{\mathsf{G}}}_{t - 1}};{{\boldsymbol{\mu}}_{{\boldsymbol{\theta}}} }\left( {{{\boldsymbol{\mathsf{G}}}_t},t} \right),\mathbf{\Sigma}_{{\boldsymbol{\theta}}} {\left( {{{\boldsymbol{\mathsf{G}}}_t},t} \right)} } \right).
\end{align}

\textit{Note that in our task, unlike the traditional GDM, the denoising model ${{{{{\boldsymbol{\varepsilon}}_{\boldsymbol{\theta }}} }}}( \cdot )$ is conditioned on side information in the form of an LR CF ${ \boldsymbol{\dot{\mathsf{G}}}}$, guiding it to progressively denoise from a Gaussian-distributed ${\boldsymbol{\mathsf{G}}_T}$ and generate the HR CF ${\boldsymbol{\mathsf{G}}_0}$. }Therefore, \eqref{eq:29} needs to be rewritten as
\begin{align}
	\label{eq:}
	{p}\left( {{{\boldsymbol{\mathsf{G}}}_{0:T}}} \left| {{ \boldsymbol{\dot{\mathsf{G}}}}} \right. \right) = {p}\left( {{{\boldsymbol{\mathsf{G}}}_T}} \right)\prod\limits_{t = 1}^T {{p_{\boldsymbol{\theta }}}\left( {{{\boldsymbol{\mathsf{G}}}_{t - 1}}\left| {{{\boldsymbol{\mathsf{G}}}_t}, { \boldsymbol{\dot{\mathsf{G}}}}} \right.} \right)},
\end{align}
and the denoising model ${{{{{\boldsymbol{\varepsilon}}_{\boldsymbol{\theta }}} }}}( \cdot )$ is trained to approximate ${p_{{\boldsymbol{\theta}}} }( {{{\boldsymbol{\mathsf{G}}}_{t - 1}}| {{{\boldsymbol{\mathsf{G}}}_t},{ \boldsymbol{\dot{\mathsf{G}}}}} })$, which is defined as
\begin{align}	
	\label{eq:}
	{p_{{\boldsymbol{\theta}}} }\!\left(\! {{{\boldsymbol{\mathsf{G}}}_{t \!-\! 1}}\!\left| {{{\boldsymbol{\mathsf{G}}}_t},\!{ \boldsymbol{\dot{\mathsf{G}}}}} \right.} \!\right) \!\!=\! \mathcal{N}\!\left(\! {{{\boldsymbol{\mathsf{G}}}_{t \!-\! 1}};\!{{\boldsymbol{\mu}}_{{\boldsymbol{\theta}}} }\left({ \boldsymbol{\dot{\mathsf{G}}}}, {{{\boldsymbol{\mathsf{G}}}_t},t} \right)\!,\!\mathbf{\Sigma}_{{\boldsymbol{\theta}}} {\left({ \boldsymbol{\dot{\mathsf{G}}}},\! {{{\boldsymbol{\mathsf{G}}}_t},\!t} \right)} } \!\right)\!\!.
\end{align}

\subsection{ELBO-Based Optimization of the CGDM}
To enable the network model ${{{{{\boldsymbol{\varepsilon}}_{\boldsymbol{\theta }}} }}}( \cdot )$ to effectively approximate the reverse process, the model parameters ${\boldsymbol{\theta}}$ need to be optimized with a specific objective. Mathematically, the latent variables ${{{\boldsymbol{\mathsf{G}}}_{1:T}}}$ and the observed sample ${{{\boldsymbol{\mathsf{G}}}_{0}}}$ \textit{conditioned} on ${ \boldsymbol{\dot{\mathsf{G}}}}$ can be represented utilizing a \textit{conditional} joint distribution $p( {{{{\boldsymbol{\mathsf{G}}}}_{0:T}}}| {{ \boldsymbol{\dot{\mathsf{G}}}}} )$. One common likelihood-based approach in generative modeling involves optimizing the model to maximize the \textit{conditional} joint probability distribution $p( {{{{\boldsymbol{\mathsf{G}}}}_{0:T}}}| {{ \boldsymbol{\dot{\mathsf{G}}}}} )$ of all observed samples. However, we only have access to the observed sample ${\boldsymbol{\mathsf{G}}_0}$ and the latent variables ${{{\boldsymbol{\mathsf{G}}}_{1:T}}}$ are unknown. Therefore, we seek to maximize the \textit{conditional} marginal distribution $p( {{{{\boldsymbol{\mathsf{G}}}}_{0}}}| {{ \boldsymbol{\dot{\mathsf{G}}}}} )$, which is given by
\begin{align}
    p\left( {{{\boldsymbol{\mathsf{G}}}_0}}\left| {{ \boldsymbol{\dot{\mathsf{G}}}}} \right. \right) = \int {p\left( {{{\boldsymbol{\mathsf{G}}}_{0:T}}}\left| {{ \boldsymbol{\dot{\mathsf{G}}}}} \right. \right)} d{{\boldsymbol{\mathsf{G}}}_{1:T}}.
\end{align}
Within the framework of variational inference, the likelihood of the observed sample ${{{\boldsymbol{\mathsf{G}}}_{0}}}$ \textit{conditioned} on ${ \boldsymbol{\dot{\mathsf{G}}}}$, known as the evidence, allows us to derive the ELBO as a proxy objective function, which can be used to optimize CGDM:
\begin{subequations}\label{eq:}
	\begin{align}
    \log p\left( {{{\boldsymbol{\mathsf{G}}}_0}}\left| {{ \boldsymbol{\dot{\mathsf{G}}}}} \right. \right) &= \log \int {p\left( {{{\boldsymbol{{\boldsymbol{\mathsf{G}}}}}_{0:T}}}\left| {{ \boldsymbol{\dot{\mathsf{G}}}}} \right. \right)} d{{\boldsymbol{{\boldsymbol{\mathsf{G}}}}}_{1:T}}\\ 
    &\mathop  \geqslant \limits^{(a)} {\mathbb{E}_{q\left( {{{\boldsymbol{\mathsf{G}}}_{1:T}}\left| {{{\boldsymbol{\mathsf{G}}}_0}} \right.} \right)}}\left( {\log \frac{{p\left( {{{\boldsymbol{\mathsf{G}}}_{0:T}}}\left| {{ \boldsymbol{\dot{\mathsf{G}}}}} \right. \right)}}{{q\left( {{{\boldsymbol{\mathsf{G}}}_{1:T}}\left| {{{\boldsymbol{\mathsf{G}}}_0}} \right.} \right)}}} \right), \label{eq:34f}
	\end{align}
\end{subequations}
where $\mathop  \geqslant \limits^{(a)} $ in \eqref{eq:34f} follows from Jensen's inequality. To further derive the ELBO, \eqref{eq:34f} can be rewritten as \eqref{eq:35e}, displayed at the top of the next page.
\begin{figure*}[ht] 
 \begin{subequations}\label{eq:}
	\begin{align}\label{eq:35}
		&\!\!\!\!\!\log p\left( {{{\boldsymbol{\mathsf{G}}}_0}}\left| {{ \boldsymbol{\dot{\mathsf{G}}}}} \right. \right)
		\geqslant {\mathbb{E}_{q\left( {{{\boldsymbol{\mathsf{G}}}_{1:T}}\left| {{{\boldsymbol{\mathsf{G}}}_0}} \right.} \right)}}\left( {\log \frac{{p\left( {{{\boldsymbol{\mathsf{G}}}_T}} \right){p_{{\boldsymbol{\theta}}}}\left( {{{\boldsymbol{\mathsf{G}}}_0}\left| {{{\boldsymbol{\mathsf{G}}}_1},{ \boldsymbol{\dot{\mathsf{G}}}} } \right.} \right)}}{{q\left( {{{\boldsymbol{\mathsf{G}}}_1}\left| {{{\boldsymbol{\mathsf{G}}}_0}} \right.} \right)}} + \log \frac{{q\left( {{{\boldsymbol{\mathsf{G}}}_1}\left| {{\boldsymbol{\mathsf{G}}_0}} \right.} \right)}}{{q\left( {{\boldsymbol{\mathsf{G}}_T}\left| {{{\boldsymbol{\mathsf{G}}}_0}} \right.} \right)}} + \log \prod\limits_{t = 2}^T {\frac{{{p_{{\boldsymbol{\theta}}}}\left( {{{\boldsymbol{\mathsf{G}}}_{t - 1}}\left| {{{\boldsymbol{\mathsf{G}}}_t},{ \boldsymbol{\dot{\mathsf{G}}}} } \right.} \right)}}{{q\left( {{{\boldsymbol{\mathsf{G}}}_{t - 1}}\left| {{{\boldsymbol{\mathsf{G}}}_t},{{\boldsymbol{\mathsf{G}}}_0}} \right.} \right)}}} } \right)\\
		&\!\!\!\!\!= \underbrace {{\mathbb{E}_{q\left( {{{\boldsymbol{\mathsf{G}}}_1}\left| {{{\boldsymbol{\mathsf{G}}}_0}} \right.} \right)}}\!\left(\! {\log {p_{{\boldsymbol{\theta}}}}\!\left(\! {{{\boldsymbol{\mathsf{G}}}_0}\!\left| {{{\boldsymbol{\mathsf{G}}}_1},\!{ \boldsymbol{\dot{\mathsf{G}}}} } \right.} \!\right)} \!\right)}_{{\mathcal{L}_{\mathbf{a}}}}\!\!-\!\! \underbrace {\sum\limits_{t = 2}^T {{\mathbb{E}_{q\left( {{{\boldsymbol{\mathsf{G}}}_t}\left| {{{\boldsymbol{\mathsf{G}}}_0}} \right.} \right)}}\!\left(\! {{D_{{\text{KL}}}}\!\left(\! {\left. {q\!\left( {{{\boldsymbol{\mathsf{G}}}_{t \!-\! 1}}\!\left| {{{\boldsymbol{\mathsf{G}}}_t},\!{{\boldsymbol{\mathsf{G}}}_0}} \right.} \!\right)} \right\|{p_{\boldsymbol{\theta }}}\!\left(\! {{{\boldsymbol{\mathsf{G}}}_{t\! - \!1}}\!\left| {{{\boldsymbol{\mathsf{G}}}_t},\!{ \boldsymbol{\dot{\mathsf{G}}}} } \right.} \!\!\right)} \!\!\right)} \!\!\right)} }_{{\mathcal{L}_{\mathbf{b}}}} \!\!-\!\! \underbrace {{D_{\text{KL}}}\!\left(\! {\left. {q\!\left(\! {{{\boldsymbol{\mathsf{G}}}_T}\!\left| {{{\boldsymbol{\mathsf{G}}}_0}} \right.}\! \right)} \right\|\!p\!\left(\! {{{\boldsymbol{\mathsf{G}}}_T}} \!\right)} \!\right)}_{{\mathcal{L}_{\mathbf{c}}}}\!\!=\!\! \mathcal{L}_{\text{ELBO}}\!\left( {\boldsymbol{\theta}}  \right) \label{eq:35e}
	\end{align}
\end{subequations}
	\hrule
\end{figure*}
Then, the parameters ${\boldsymbol{\theta}}$ of CGDM, can be learned by maximizing the ELBO:
\begin{align}\label{eq:36}
	\mathop {{\text{argmin}}}\limits_{\boldsymbol{\theta}} 
	\mathcal{L}\left( {{\boldsymbol{\theta}}} \right) = {\mathbb{E} }\left( { - {\mathcal{L}_{{\text{ELBO}}}}\left( {{\boldsymbol{\theta}}} \right)} \right) = {\mathbb{E}}\left( {{\mathcal{L}_{\mathbf{c}}} + {\mathcal{L}_{{\mathbf{b}}}} - {\mathcal{L}_{\mathbf{a}}}} \right),
\end{align}
where $\mathcal{L}( \boldsymbol{{\theta}})$ is the objective function for CGDM training, ${\mathcal{L}_{\mathbf{c}}}$ is constant from \eqref{eq:35e} and can be excluded from the optimization, and ${\mathcal{L}_{\mathbf{a}}}$ can be approximated and optimized via a Monte Carlo estimate \cite{Ho11}. From the above analysis, it is evident that the training objective of CGDM is primarily determined by $\mathcal{L}_{{\mathbf{b}}}$. It can be found that the training objective of CGDM is to approximate the transition density in the reverse process as closely as possible, thereby minimizing the Kullback-Leibler (KL) divergence between ${p_{{\boldsymbol{\theta}}} }\!\left(\! {{{\boldsymbol{\mathsf{G}}}_{t - 1}}\!\left| {{{\boldsymbol{\mathsf{G}}}_t},{ \boldsymbol{\dot{\mathsf{G}}}}} \right.} \!\right) = \mathcal{N}\left(\! {{{\boldsymbol{\mathsf{G}}}_{t - 1}};{{\boldsymbol{\mu}}_{{\boldsymbol{\theta}}} }\left({ \boldsymbol{\dot{\mathsf{G}}}}, {{{\boldsymbol{\mathsf{G}}}_t},t} \right),\mathbf{\Sigma}_{{\boldsymbol{\theta}}} {\left({ \boldsymbol{\dot{\mathsf{G}}}}, {{{\boldsymbol{\mathsf{G}}}_t},t} \right)} } \!\right)$ and $q\left( {{{\boldsymbol{\mathsf{G}}}_{t - 1}}\left| {{{\boldsymbol{\mathsf{G}}}_t},{{\boldsymbol{\mathsf{G}}}_0}} \right.} \right) = \mathcal{N}\left( {{{\boldsymbol{\mathsf{G}}}_{t - 1}};{{{{\tilde {\boldsymbol{\mu}}_t} }}}\left( {{{\boldsymbol{\mathsf{G}}}_t},{{\boldsymbol{\mathsf{G}}}_0}} \right),{{\tilde \beta }_t}{\mathbf{I}}} \right)$.

Therefore, we need to obtain explicit expressions for the mean ${{{{\tilde {\boldsymbol{\mu}}_t} }}}( {{{\boldsymbol{\mathsf{G}}}_t},{{\boldsymbol{\mathsf{G}}}_0}})$ and variance ${{\tilde \beta }_t}{\mathbf{I}}$ of $q( {{{\boldsymbol{\mathsf{G}}}_{t - 1}}| {{{\boldsymbol{\mathsf{G}}}_t},{{\boldsymbol{\mathsf{G}}}_0}} } )$. Specifically, $q( {{{\boldsymbol{\mathsf{G}}}_{t - 1}}| {{{\boldsymbol{\mathsf{G}}}_t},{{\boldsymbol{\mathsf{G}}}_0}} } )$ can be expressed as
\begin{align}\label{eq:qt-1}
	q\left( {{{{\boldsymbol{\mathsf{G}}}}_{t - 1}}\left| {{{{\boldsymbol{\mathsf{G}}}}_t},{{{\boldsymbol{\mathsf{G}}}}_0}} \right.} \right) 
	= \frac{{q\left( {{{{\boldsymbol{\mathsf{G}}}}_t}|{{{\boldsymbol{\mathsf{G}}}}_{t - 1}},{{{\boldsymbol{\mathsf{G}}}}_0}} \right)q\left( {{{{\boldsymbol{\mathsf{G}}}}_{t - 1}}|{{{\boldsymbol{\mathsf{G}}}}_0}} \right)}}{{q\left( {{{{\boldsymbol{\mathsf{G}}}}_t}|{{{\boldsymbol{\mathsf{G}}}}_0}} \right)}}.
\end{align}
Combining \eqref{eq:ht} and \eqref{eq:chmd}, \eqref{eq:qt-1} can be further rewritten as \eqref{eq:38e}, displayed at the top of this page.
\begin{figure*}[ht] 
	\begin{align}  \label{eq:38e}
		q\left( {{{\boldsymbol{\mathsf{G}}}_{t - 1}}\left| {{{{\boldsymbol{\mathsf{G}}}}_t},{{{\boldsymbol{\mathsf{G}}}}_0}} \right.} \right) 
	    = \exp  \left( { - \frac{1}{2}\left( {\frac{1}{{\frac{{\left( {1 - {\alpha _t}} \right)\left( {1 - {{\bar \alpha }_{t - 1}}} \right)}}{{1 - {{\bar \alpha }_t}}}}}} \right)\left( {{\boldsymbol{\mathsf{G}}}_{t - 1}^2 - 2\frac{{\sqrt {{\alpha _t}} \left( {1 - {{\bar \alpha }_{t - 1}}} \right){{\boldsymbol{\mathsf{G}}}_t} + \sqrt {{{\bar \alpha }_{t - 1}}} \left( {1 - {\alpha _t}} \right){{\boldsymbol{\mathsf{G}}}_0}}}{{1 - {{\bar \alpha }_t}}}{{\boldsymbol{\mathsf{G}}}_{t - 1}}} \right)} \right) 
	\end{align}
	\hrule
\end{figure*}
Based on \eqref{eq:38e}, the mean and variance of $q( {{{\boldsymbol{\mathsf{G}}}_{t - 1}}| {{{\boldsymbol{\mathsf{G}}}_t},{{\boldsymbol{\mathsf{G}}}_0}} } )$ are explicitly expressed as:
\begin{align}
	\label{eq:41}
	{{{{\tilde {\boldsymbol{\mu}}_t} }}}\left( {{{\boldsymbol{\mathsf{G}}}_t},{{\boldsymbol{\mathsf{G}}}_0}} \right) &= \frac{1}{{\sqrt {{\alpha _t}} }}\left({{{\boldsymbol{\mathsf{G}}}}_t} - \frac{{{\beta _t}}}{{\sqrt {1 - {{\bar \alpha }_t}} }}{\boldsymbol{\varepsilon}_t}\right),\\
	\label{eq:}
	{{\tilde \beta }_t} &= \frac{{1 - {{\bar \alpha }_{t - 1}}}}{{1 - {{\bar \alpha }_t}}}{\beta _t}.
\end{align}
Generally, the variance $\mathbf{\Sigma}_{{\boldsymbol{\theta}}} {({ \boldsymbol{\dot{\mathsf{G}}}}, {{{\boldsymbol{\mathsf{G}}}_t},t} )}$ is set as a constant ${{\tilde \beta }_t}{\mathbf{I}}$ \cite{Ho11,jinglobecom}. Therefore, to ensure that the denoising transition density closely approximates the ground-truth denoising transition density, we can simplify the optimization of the KL divergence term to minimizing the difference between the expectations of the above two distributions. In this case, we only need to train CGDM to predict ${{{{\tilde {\boldsymbol{\mu}}_t} }}}( {{{\boldsymbol{\mathsf{G}}}_t},{{\boldsymbol{\mathsf{G}}}_0}})$:
\begin{align}\label{eq:42}
	{{\boldsymbol{\mu}}_{\boldsymbol{\theta }}}\left( {{ \boldsymbol{\dot{\mathsf{G}}}}, {{\boldsymbol{\mathsf{G}}}_t},t} \right) \!=\! \frac{1}{{\sqrt {{\alpha _t}} }}\left( {{{\boldsymbol{\mathsf{G}}}_t} - \frac{{1 - {\alpha _t}}}{{\sqrt {1 - {{\bar \alpha }_t}} }}{{\boldsymbol{\varepsilon}}_{\boldsymbol{\theta }}}\left( {{ \boldsymbol{\dot{\mathsf{G}}}}, {{\boldsymbol{\mathsf{G}}}_t},t} \right)} \right),
\end{align}

Recall that the KL divergence is defined as 
\begin{align}\label{eq:kl}
	{D_{\rm{KL}}}&\left( {\mathcal{N}\left( {\bf{x};{{\boldsymbol{\mu }}_{\bf{x}}},{{\mathbf{\Sigma }}_{\bf{x}}}} \right)\left\| {\mathcal{N}\left( {{\bf{y}};{{\boldsymbol{\mu }}_{\bf{y}}},{{\mathbf{\Sigma }}_{\bf{y}}}} \right)} \right.} \right) = \frac{1}{2}\Big[\log \frac{{\left| {{{\mathbf{\Sigma }}_{\bf{y}}}} \right|}}{{\left| {{{\mathbf{\Sigma }}_{\bf{x}}}} \right|}} - d \nonumber\\
	&+ {\rm{tr}}\left( {{\mathbf{\Sigma }}_{\bf{y}}^{ - 1}{{\mathbf{\Sigma }}_{\bf{x}}}} \right) + {\left( {{{\boldsymbol{\mu }}_{\bf{y}}} - {{\boldsymbol{\mu }}_{\bf{x}}}} \right)^T}{\mathbf{\Sigma }}_{\bf{y}}^{ - 1}\left( {{{\boldsymbol{\mu }}_{\bf{y}}} - {{\boldsymbol{\mu }}_{\bf{x}}}} \right)\Big],
\end{align}
where $d$ represents the dimension of $\bf{x}$. Substituting \eqref{eq:41}, \eqref{eq:42}, and \eqref{eq:kl} into $\mathcal{L}_{{\mathbf{b}}}$ in \eqref{eq:35e}, yields
\begin{align}\label{eq:45}
  \!\mathcal{L}_{{\mathbf{b}}} 
  \!\!=\!\! \prod\limits_{t = 2}^T  \!{\mathbb{E}_{q\left( {{{\boldsymbol{\mathsf{G}}}_t}\left| {{{\boldsymbol{\mathsf{G}}}_0}} \right.} \right)}}\!\!\left(\! {\frac{{{{\left(\! {1 \!-\! {\alpha _t}} \!\right)}^2}}}{{2\tilde \beta _t^2{\alpha _t}\left(\! {1 \!-\! {{\bar \alpha }_t}} \!\right)}}\!\left\| {{{\boldsymbol{\varepsilon }}_t} \!-\! {{\boldsymbol{\varepsilon }}_{\boldsymbol{\theta }}}\left(\! {{ \boldsymbol{\dot{\mathsf{G}}}} ,{{\boldsymbol{\mathsf{G}}}_t},t} \!\right)} \right\|_2^2} \!\right)\!.
\end{align}
Substituting \eqref{eq:35e} and \eqref{eq:45} into \eqref{eq:36}, the objective function $\mathcal{L}( \boldsymbol{{\theta}})$ for the CGDM can be further simplified to
\begin{align}\label{eq:46} 
	\!\!\!\!\! {\underline {\mathcal{L}}} (\boldsymbol{\theta}) \!\!&:=\!\!\! \sum\limits_{t = 1}^T\!{\mathbb{E}_{{{{{\boldsymbol{\mathsf{G}}}_0}}}\!,{{\boldsymbol{\varepsilon }}_t} }}\!\Big( \!\big\| {{\boldsymbol{\varepsilon }}_t} \!\!-\!\! {{\boldsymbol{\varepsilon }}_{\boldsymbol{\theta }}}\big({{ \boldsymbol{\dot{\mathsf{G}}}}},\! \underbrace{\sqrt {{{\bar \alpha }_t}} {{{\boldsymbol{\mathsf{G}}}_0}} \!+\! \sqrt {1 \!-\! {{\bar \alpha }_t}} {{\boldsymbol{\varepsilon }}}}_{{{\boldsymbol{\mathsf{G}}}_t}},\!t \big) \big\|_2^2 \Big)\!.\!\!\!\!\!\!\!
\end{align}
Based on the trained CGDM, given any noise-contaminated CF ${{\boldsymbol{\mathsf{G}}}_t}$, the trained CGDM can leverage the side information in ${{ \boldsymbol{\dot{\mathsf{G}}}}}$ to predict the noise ${{\boldsymbol{\varepsilon }}_t}$ and subsequently obtain an approximation of the target CF ${{\hat{{\boldsymbol{\mathsf{G}}}}_0}}$ through transformation \eqref{eq:ht}, i.e., 
\begin{align}\label{eq:47}
	{{\hat{{\boldsymbol{\mathsf{G}}}}_0}} \!=\! \frac{1}{{\sqrt {{{\bar \alpha }_t}} }}\!\!\left(\!\! {{{{\boldsymbol{\mathsf{G}}}_t}} \!-\! \sqrt {1 \!-\! {{\bar \alpha }_t}} {{\boldsymbol{\varepsilon }}_{\boldsymbol{\theta }}}\big({{ \boldsymbol{\dot{\mathsf{G}}}}}, \underbrace{\sqrt {{{\bar \alpha }_t}} {{{\boldsymbol{\mathsf{G}}}_0}} \!+\! \sqrt {1 \!-\! {{\bar \alpha }_t}} {{\boldsymbol{\varepsilon }}}}_{{{\boldsymbol{\mathsf{G}}}_t}},t \big)} \!\!\right)\!.
\end{align}
Through reparameterization, \eqref{eq:47} represents the results of iterative refinements, with each iteration in our proposed CGDM being represented by 
\begin{align}\label{eq:48}
	\!\!\!\!\!{{{\boldsymbol{\mathsf{G}}}_{t\!-\!1}}} \!\leftarrow\! \frac{1}{{\sqrt {{\alpha _t}} }}\!\!\left(\!\! {{{{\boldsymbol{\mathsf{G}}}}_t} \!\!-\!\! \frac{{1 \!-\! {\alpha _t}}}{{\sqrt {1 \!-\! {{\bar \alpha }_t}} }}{{\boldsymbol{\varepsilon}}_{\boldsymbol{\theta }}}\!\left(\!{{ \boldsymbol{\dot{\mathsf{G}}}}, {{\boldsymbol{\mathsf{G}}}_t},t} \!\right)} \!\!\right) \!+\! \sqrt { \frac{{1 \!-\! {{\bar \alpha }_{t \!-\! 1}}}}{{1 \!-\! {{\bar \alpha }_t}}}{\beta _t}} {\boldsymbol{\varepsilon}^*},\!\!\!\!\!\!
\end{align}
where ${\boldsymbol{\varepsilon}^*} \sim\mathcal{N}( {\mathbf{0},\mathbf{I}})$. Note that the noise estimation step in \eqref{eq:48} is analogous to a step in Langevin dynamics within score-based generative models \cite{song2019generative}, which is equivalent to the estimation of the first derivative of the log-likelihood of the observed samples, also known as the gradient or Stein score. For clarity, we summarize the training process and iterative inference process of the proposed CGDM in \textbf{\alref{alg:train}} and \textbf{\alref{alg:infer}}, respectively. 

\begin{figure*}[!t]
	\centering
	\includegraphics[width=0.966\textwidth]{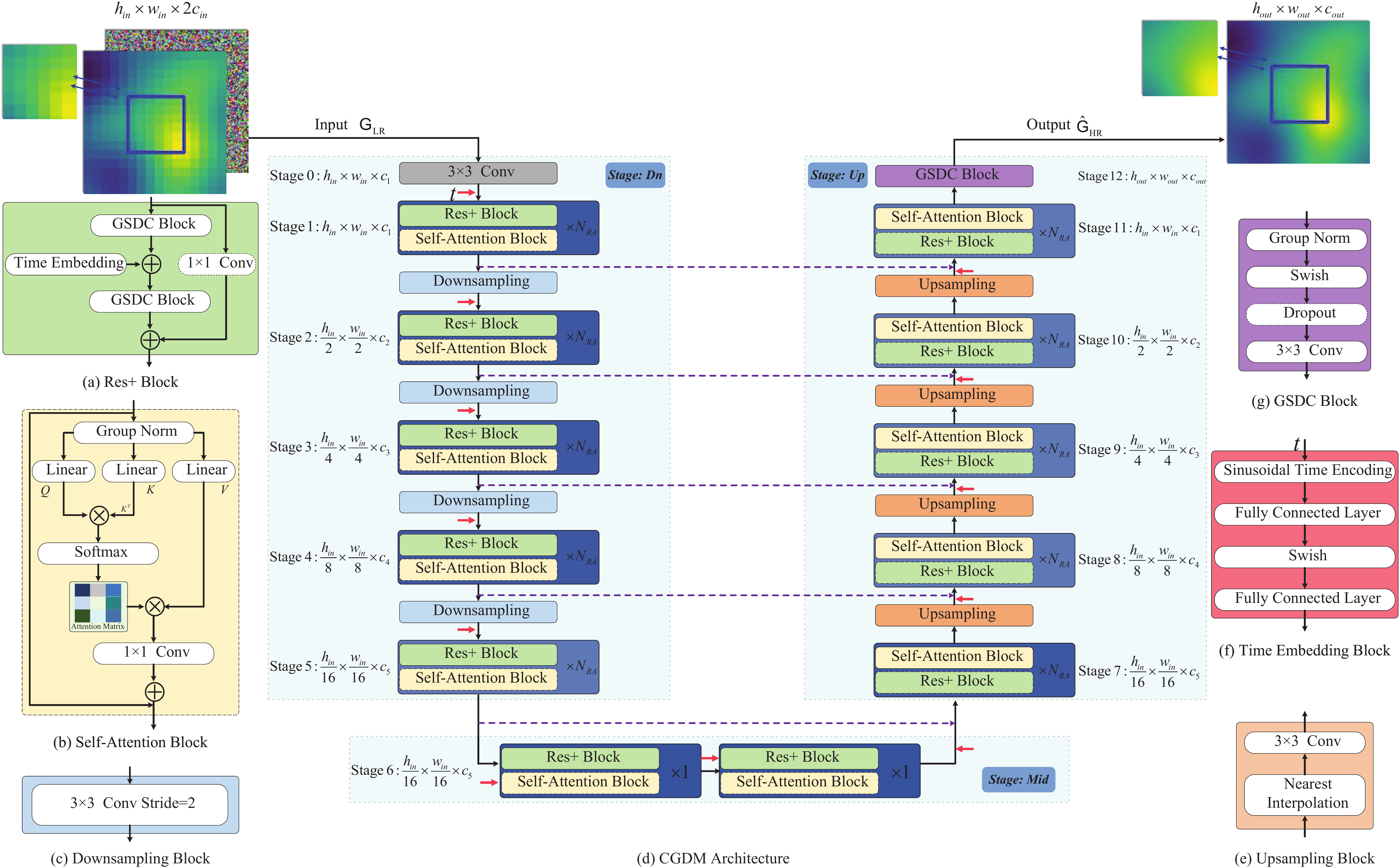}
	\captionsetup{font=footnotesize}
	\caption{Diagram and key modules of the CGDM architecture. Specifically, the network architecture of CGDM consists of three primary stages: \textit{Dn} (substages 1-5), \textit{Mid} (substage 6), and \textit{Up} (substages 7-12). The blocks included in each stage are illustrated in the subfigures (a), (b), (c), (e), (f), and (g). (d) shows the network architecture of the proposed CGDM. Additionally, the red and purple arrows represent the embedding of the time constant $t$ and the skip connections, respectively. Taking the $\times4$ HR CF reconstruction task as an example, the LR CF with a size of $32\times32\times3$ is upsampled to the target resolution (i.e., $128\times128\times3$), and concatenated with noise of the same resolution along the channel dimension to form the input, resulting in a size of $128\times128\times6$. The number $2c_{in}$ of input channels is expanded to $c_1$, representing the number of base channels in the latent space, after passing through substage 0. Note that within the same substage, the height $h$ and width $w$ of the feature maps remain unchanged, while the number $c$ of channels at different substages is controlled by the channel number multiplier $\bar \eta  = {c_1}:{c_2}:{c_3}:{c_4}:{c_5}$. The specific values of $c_1$, $\bar \eta $, and $N_{\rm{RA}}$ will be discussed in Subsection \ref{Experiment Results}. }
	\label{fig:model}
\end{figure*}

​



\subsection{Network Architecture of the Proposed CGDM}\label{subsec:net}
In this subsection, we present the architecture of CGDM, a variant of the U-Net model, as illustrated in \figref{fig:model}. The network is composed of three primary stages: Dn, Mid, and Up. For clarity, we provide a concise overview of the key components in each stage, including the time embedding block, Res$^+$ block, self-attention block, downsampling block, and upsampling block.

\textit{1) Time Embedding Block:} 
In order to encode the time parameter $[1,..., t,...,T]$ in the diffusion process, sine and cosine functions with different frequencies are employed, akin to the sinusoidal positional encoding approach used in \cite{vaswani2017attention}. This approach results in the time embedding vector ${{\mathbf{\Gamma }}}_t$, which effectively captures the temporal characteristics. Let ${{{\Gamma }}}^j_t \in {\mathbb{R}}$ denote the $j$-th component of the time embedding vector at time step $t$, defined as 
\begin{align}\label{eq：}
	{{\Gamma}}_t^{(j)} = \left\{ \begin{gathered}
		\cos \left( {\frac{t}{{{{10000}^{{{2j} \mathord{\left/
									{\vphantom {{2j} {{c_{{\rm{time}}}}}}} \right.
									\kern-\nulldelimiterspace} {{c_{{\rm{time}}}}}}}}}}} \right),\text{ if}\quad j \quad \text{is odd}\hfill \\ 
	   \sin \left( {\frac{t}{{{{10000}^{{{2j} \mathord{\left/
	   							{\vphantom {{2j} {{c_{{\rm{time}}}}}}} \right.
	   							\kern-\nulldelimiterspace} {{c_{{\rm{time}}}}}}}}}}} \right),\text{ if}\quad j \quad \text{is even} \hfill \\
	\end{gathered}  \right.,
\end{align}
where $j = 0,1,2,...,{c_{{\rm{time}}}}/2-1$, $c_{{\rm{time}}}$ is the dimension of the time embedding vector. Then, the embedding vector for the time constant $t$ can be expressed as
\begin{align}\label{eq:50}
	{{\mathbf{\Gamma }}}_t &= \Big[\sin \big( {{w_0}t} \big),\cos \big( {{w_0}t} \big),\sin \big( {{w_1}t} \big),\cos \big( {{w_1}t} \big),\nonumber \\
	&\qquad\qquad \ldots,\sin \big( {{w_{\tfrac{{{c_{{\rm{time}}}}}}{2} - 1}}t} \big),\cos \big( {{w_{\tfrac{{{c_{{\rm{time}}}}}}{2} - 1}}t} \big)\Big],
\end{align}
where ${w_j} = {1 \mathord{\left/
		{\vphantom {1 {{{10000}^{{{2j} \mathord{\left/
									{\vphantom {{2j} {{c_{{\text{time}}}}}}} \right.
									\kern-\nulldelimiterspace} {{c_{{\rm{time}}}}}}}}}}} \right.
		\kern-\nulldelimiterspace} {{{10000}^{{{2j} \mathord{\left/
						{\vphantom {{2j} {{c_{{\rm{time}}}}}}} \right.
						\kern-\nulldelimiterspace} {{c_{{\text{time}}}}}}}}}}$. Leveraging \eqref{eq:50}, the embedding vectors for any given time ${{\mathbf{\Gamma }}}_{t + \Delta t}$ can be obtained through a linear transformation expressed as					
\begin{align}\label{eq：51}
	{{\mathbf{\Gamma }}}_{t + \Delta t}^T = \left[ {\begin{array}{*{20}{c}}
			{\sin \Big( {{w_0}\big( {t + \Delta t} \big)} \Big)} \\ 
			{\cos \Big( {{w_0}\big( {t + \Delta t} \big)} \Big)} \\ 
			...\\
			{\sin \Big( {{w_{\tfrac{{{c_{{\rm{time}}}}}}{2} - 1}}\big( {t + \Delta t} \big)} \Big)} \\
			{\cos \Big( {{w_{\tfrac{{{c_{{\rm{time}}}}}}{2} - 1}}\big( {t + \Delta t} \big)} \Big)} 
	\end{array}} \right] = {{\bf{M}}_{\Delta t}} \cdot {{\mathbf{\Gamma }}}_t^T.
\end{align}
${\bf{M}}_{\Delta t}$ is a linear transform matrix defined as 
\begin{align}
	\mathbf{M}_{\Delta t}=\begin{pmatrix}\mathbf{R}(w_0\Delta t)&\cdots&0\\\vdots&\ddots&\vdots\\0&\cdots&\mathbf{R}\left(w_{\frac{c_{\mathrm{time}}}{2}-1}\Delta t\right)\end{pmatrix},
\end{align}
where $\mathbf{R}(w_0\Delta t)=\begin{bmatrix}\cos(w_0\Delta t)&&\sin(w_0\Delta t)\\-\sin(w_0\Delta t)&&\cos(w_0\Delta t)\end{bmatrix}$. To enable the model to capture more intricate temporal features, the time embedding is further enhanced through two fully connected layers and an activation layer, expressed as ${{{{{\mathbf{\Gamma }}'}}}_t} = {f_{{\rm{Ful}}}}( {{f_{{\rm{Swi}}}}( {{f_{{\rm{Ful}}}}( {{{\mathbf{\Gamma }}}_t} )} )} )$, where ${f_{{\rm{Ful}}}}(  \cdot  )$ and ${f_{{\rm{Swi}}}}(  \cdot  )$ represent the fully connected layer and swish activation function layer, respectively. The components of the time embedding block are summarized in \figref{fig:model} (f).

\textit{2) Res$^+$ Block:} We design deeper feature extraction networks aimed at learning higher-level semantic information from the CF. Nonetheless, deep neural networks frequently experience model degradation as a result of challenges like vanishing and exploding gradients. To this end, we introduce residual connections as a means of alleviating these concerns. Specifically, let ${\bf{X}} \in {\mathbb{R}^{{h_{{\rm{in}}}} \times {w_{{\rm{in}}}} \times {c_{{\rm{in}}}}}}$, ${{\mathbf{\Gamma }}}_t$ be the feature map and time embedding vector inputs, respectively. The output ${\bf{X}}'\in {\mathbb{R}^{{h_{{\rm{in}}}} \times {w_{{\rm{in}}}} \times {c_{{\rm{out}}}}}}$ of Res$^+$ block is given by
\begin{align}\label{eq：}
	{\bf{X}}' = \left\{ \begin{gathered}
		\boldsymbol{\Omega }  + {{f}_{{\rm{Conv}},1}}\left( {\bf{X}} \right),\text{ if }{c_{{\rm{out}}}} \ne {c_{{\rm{in}}}} \hfill \\
		\boldsymbol{\Omega }  + {\bf{X}},\text{ else} \hfill \\ 
	\end{gathered}  \right.,
\end{align}
where ${f_{{\rm{Conv}},1}}( \cdot) $ denotes $1\times 1$ convolution operation, and $\boldsymbol{\Omega }  \in {\mathbb{R}^{{h_{{\rm{in}}}} \times {w_{{\rm{in}}}} \times {c_{{\rm{out}}}}}}$ is defined as
\begin{align}
  {\boldsymbol{\Omega}}  = g\left( {g({\mathbf{X}}) + {f_{{\text{Ful}}}}\left( {{f_{{\text{Swi}}}}\left( {{f_{{\text{Ful}}}}\left( {{{\mathbf{{{\Gamma }}}}_t}} \right)} \right)} \right)} \right).
\end{align}
Note that $g({\mathbf{X}}) = {f_{{\rm{Conv}},3}}( {{f_{{\rm{Dro}}}}( {{f_{{\text{Swi}}}}( {{f_{{\rm{Gn}}}}( {\mathbf{X}} )} )} )} )$, where ${f_{{\rm{Conv}},3}}(\cdot ) $ is $3\times3$ convolution operation. ${f_{{\rm{Dro}}}}(  \cdot )$, ${f_{{\rm{Gn}}}}(  \cdot )$ are the dropout and group normalization layers, respectively. The components of the Res$^+$ block are summarized in \figref{fig:model} (a).

\textit{3) Self-Attention Block:} To capture global structural information, we introduce a self-attention mechanism to improve the interaction between global and local features. Specifically, to implement the self-attention mechanism, a normalized attention matrix is introduced to represent varying degrees of attention to the input. Greater weights are assigned to more significant input components. The final output is generated by weighting the input according to the attention weights indicated in the attention matrix. Specifically, given the input $\mathbf{Z}\! =\! {[{\mathbf{z}_1},...,{\mathbf{z}_m}]^T} \in {\mathbb{R}^{{d_m} \times {d_n}}}$, ${{d_m}}$ denotes the number of image patches and ${{d_n}}$ is the feature dimension of each patch, three different linear transformations are applied to ${\mathbf{z}_j}$ \cite{9613212,vaswani2017attention}:
\begin{subequations}\label{eq:spdb}
	\begin{align}
		{\mathbf{k}_j} &= {\mathbf{z}_j}{\mathbf{W}^k},\quad j = 1,...,{d_n},\label{eq:dbso}\\
		{\mathbf{q}_j} &= {\mathbf{z}_j}{\mathbf{W}^q},\quad j = 1,...,{d_n},\label{eq:dbsc}\\
		{\mathbf{v}_j} &= {\mathbf{z}_j}{\mathbf{W}^v},\quad j = 1,...,{d_n},\label{eq:dbss} 
	\end{align}
\end{subequations}
where ${\mathbf{k}_j} \in {\mathbb{R}^{1 \times {d_k}}}$, ${\mathbf{q}_j} \in {\mathbb{R}^{1 \times {d_q}}}$, and ${\mathbf{v}_j} \in {\mathbb{R}^{1 \times {d_n}}}$ are the key, query, and value vector, respectively. ${\mathbf{W}^k} \in {\mathbb{R}^{{d_n} \times {d_k}}}$, ${\mathbf{W}^q} \in {\mathbb{R}^{{d_n} \times {d_q}}}$, and ${\mathbf{W}^v} \in {\mathbb{R}^{{d_n}\times {d_n}}}$ represent the respective trainable transformation matrices, with ${d_k} = {d_q}$. In detail, the weight allocation function is determined by ${\mathbf{k}_j}$ and ${\mathbf{q}_{j'}}$. A higher correlation ${\mathbf{q}_{j'}}\mathbf{k}_j^T$ implies that the features of the $j$-th input patch ${\mathbf{z}_j}$ hold greater importance for the ${j'}$-th output patch. Generally, this correlation can be adaptively adjusted based on the input $\mathbf{Z}$ and the matrices ${\mathbf{W}^k}$ and ${\mathbf{W}^q}$. For clarity, the matrix forms of \eqref{eq:dbso}-\eqref{eq:dbss} are presented as ${\mathbf{K}} = {\mathbf{Z}}{\mathbf{W}^k}$, ${\mathbf{Q}} = {\mathbf{Z}}{\mathbf{W}^q}$, ${\mathbf{V}} = {\mathbf{Z}}{\mathbf{W}^v}$,
where $\mathbf{K} = {[{\mathbf{k}_1},...,{\mathbf{k}_m}]^T} \in {\mathbb{R}^{{d_m} \times {d_k}}}$, $\mathbf{Q} = {[{\mathbf{q}_1},...,{\mathbf{q}_m}]^T} \in {\mathbb{R}^{{d_m} \times {d_q}}}$ and $\mathbf{V} = {[{\mathbf{v}_1},...,{\mathbf{v}_m}]^T} \in {\mathbb{R}^{{d_m} \times {d_n}}}$.

Leveraging $\mathbf{K}$ and $\mathbf{Q}$, we can obtain the attention matrix $\mathbf{S}_{\rm{AM}}  \in {\mathbb{R}^{{d_m} \times {d_m}}}$, which is denoted as 
\begin{align}\label{eq:cchmd}
	\mathbf{S}_{\rm{AM}}  = \rm{Softmax} \left( {\frac{{\mathbf{Q}{\mathbf{K}^\textit{T}}}}{{\mathit{\sqrt \eta} }}} \right),
\end{align}
where $\rm{Softmax} ( \mathbf{Z})  = {\exp ( \mathit{{z_j}}) }/{{\sum {\exp ( \mathit{{z_j}}) } }}$, and $\sqrt \eta >0$ is a scaling factor. Each column of the attention matrix is a vector of attention scores, i.e., each score is a probability, where all scores are non-negative and sum up to 1. Note that when the key vector $\mathbf{K}[j,:]$ and the query $\mathbf{Q}[j',:]$ has a better match, the corresponding attention score $\mathbf{S}_{\rm{AM}} [j',j]$ is higher. Thus, the output of the attention mechanism corresponding to the $r$-th component can be represented by the weighted sum of all inputs, i.e., ${\mathbf{z}'_r} = \sum\nolimits_j {\mathbf{S}_{\rm{AM}}  [r,j]{\mathbf{v}_j}}  = \mathbf{S}_{\rm{AM}}  [r,:] \cdot \mathbf{V}$, where ${\mathbf{z}'_r} \in {\mathbb{R}^{1 \times {d_n}}}$ represents the $r$-th output, which is computed by adaptively focusing on the inputs based on the attention score $\mathbf{S}_{\rm{AM}}[r,j]$. When the attention score $\mathbf{S}_{\rm{AM}}[r,j]$ is higher, the associated value vector ${\mathbf{v}_j}$ will have a more significant impact on the $r$-th output patch. Finally, the output ${\bf{Z}}'$ of the attention block is given by 
\begin{align}\label{eq:cmd}
	{\bf{Z}}' \!=\!\mathbf{S}_{\rm{AM}}[:,:]\cdot \mathbf{V}\!=\!\rm{Softmax}\mathit{\left( {\frac{{\mathbf{Z}{\mathbf{W}^q}{\mathbf{W}^{{k^T}}}{\mathbf{Z}^T}}}{{\sqrt \eta }}} \right)\mathbf{Z}{\mathbf{W}^v}},
\end{align}
where $\mathbf{O} = {[{\mathbf{o}_{1,...,}}{\mathbf{o}_m}]^T} \in {\mathbb{R}^{{d_m} \times {d_n}}}$. The components of the self-attention block are summarized in \figref{fig:model} (b). Notably, as illustrated in \figref{fig:model}, the network framework of the proposed CGDM consists of three primary stages: Dn, Mid, and Up. Each sub-stage within these stages comprises $N_{\rm{RA}}$ Res$^+$ blocks, along with their corresponding self-attention blocks.

\section{Lightweight CGDM for SR CF}\label{sec:}
Considering that the proposed CGDM trades off significant memory consumption and latency for outstanding performance, its deployment on personal computers and even mobile devices is highly constrained. Specifically, CGDM utilizes a variant of U-Net as their backbone framework, with most of the memory consumption and latency arising from this architecture. To this end, in this section, we leverage the additive property of network layers and design a one-shot pruning approach along with the corresponding knowledge distillation technique to obtain a lightweight CGDM (LiCGDM). 

\begin{algorithm}[!b]
	\caption{Efficient Layer Pruning for LiCGDM}
	\label{alg:algtrhpf}
	\begin{algorithmic}[1]
		\STATE \textbf{Input:} A teacher (original) network with prunable layers ${\tilde L^{\tilde n}} = \left\{ {{{\tilde l}_1},...,{{\tilde l}_{\tilde n}}} \right\}$, a CF dataset $D$, the number of parameters $\mathcal{P}$ to be pruned
		\STATE \textbf{Initialize:} Arrays $values [\text{ }]=0$ and $weights [\text{ }]=0$, Knapsack capacity $\mathcal{C}=0$
		\FOR{${p_i}$ in ${\tilde L^{\tilde n}}$ }
		\STATE
		Calculate the value of the objective function \eqref{eq:61a} on $D$, $values [i]=\mathbb{E}\left\|  {{{\boldsymbol{\varepsilon }}_{{\text{tea}}}} - {\boldsymbol{\varepsilon }}\left( {{p_i}} \right)} \right\|_2^2$ 
		\STATE
		Calculate the number of parameters at the $p_i$-th layer, $weights [i]={\rm{Params}}\left(p_i \right) $
		\ENDFOR
		\STATE
		$\mathcal{C}=\mathcal{P}$
		\STATE
		Obtain $P^{\tilde m}$ by solving problem \eqref{eq:61} using the dynamic programming algorithm, given $values$, $weights$, $\mathcal{C}$.
		\STATE \textbf{Return} $P^{\tilde m}$
		\STATE \textbf{Output:} The set of layers to be pruned $P^{\tilde m} \in {\tilde L^{\tilde n}} $
	\end{algorithmic}
\end{algorithm}
\subsection{Efficient Layer Pruning for CGDM}
Given a specific pruning ratio, the objective of layer pruning is to eliminate a subset of prunable layers from the U-Net architecture in CGDM while minimizing the degradation in the model's performance. We define the set of all the ${\tilde n}$ prunable layers as ${\tilde L^{\tilde n}} = \{ {{{\tilde l}_1},...,{{\tilde l}_{\tilde n}}} \}$. Inspired by \cite{zhang2024laptop}, we also minimize the mean-squared-error (MSE) loss between the final output of the original network (referred to as the teacher) and that of the pruned network (referred to as the student) as the pruning objective. Specifically, let ${{{{{\boldsymbol{\varepsilon}}_{{\rm{tea}}}} }}}$ define the output of the original CGDM, $p_i \in {\tilde L^{\tilde n}} $ represent the $p$-th pruned layer, and ${{{{{\boldsymbol{\varepsilon}}} ( p_1,p_2,...,p_{\tilde m})  }}}$ denote the output of the network with layers $p_1$,$p_2$,..., $p_{\tilde m}$ pruned, where ${\tilde m}$ is an uncertain variable. Thus, this optimization problem can be formulated as
\begin{subequations}\label{eq:57}
	\begin{align}
		&\quad\mathop {\min }\limits_{{p_1},{p_2},...,{p_{\tilde m}}} \mathbb{E}\left\| {{{{{{\boldsymbol{\varepsilon}}_{{\rm{tea}}}} }}} - {{{{{\boldsymbol{\varepsilon}}} \left( p_1,p_2,...,p_{\tilde m}\right)  }}}} \right\|_2^2,\label{eq:57a}\\
		&{\rm{s.t.}}\left\lbrace {{p_1},{p_2},...,{p_{\tilde m}}} \right\rbrace  \subset {{\tilde L}^{\tilde n}},\sum\limits_{i = 1}^{\tilde m} {\rm{Params}} \left( {{p_i}} \right) \geq \mathcal{P},\label{eq:57b}
	\end{align}
\end{subequations}
where ${\rm{Params}} \left( {p_i} \right)$ represents the number of parameters in the $p_i$-th layer. Note that $\mathcal{P}$ denotes the number of parameters to be pruned, calculated by multiplying the total number of parameters in the teacher network by the pruning ratio. Solving the optimization problems \eqref{eq:57} is NP-hard; therefore, we need to find a surrogate objective. Capitalizing on the triangle inequality, we can obtain the upper bound of \eqref{eq:57a}:
\begin{align}\label{eq:}
	&\quad\mathbb{E}\left\| {{{\boldsymbol{\varepsilon }}_{{\text{tea}}}} - {\boldsymbol{\varepsilon }}\left( {{p_1},{p_2},...,{p_{\tilde m}}} \right)} \right\|_2^2 \leq \mathcal{L}_{{\rm{upper}}},
\end{align}
where $\mathcal{L}_{{\rm{upper}}}\!=\! \mathbb{E}\left\| {{{\boldsymbol{\varepsilon }}_{{\text{tea}}}} - {\boldsymbol{\varepsilon }}\left( {{p_1}} \right)} \right\|_2^2 + \mathbb{E}\left\| {{\boldsymbol{\varepsilon }}\left( {{p_1}} \right) - {\boldsymbol{\varepsilon }}\left( {{p_1},{p_2}} \right)} \right\|_2^2 + ... + \mathbb{E}\left\| {{\boldsymbol{\varepsilon }}\left( {{p_1},{p_2},...,{p_{\tilde m - 1}}} \right) - {\boldsymbol{\varepsilon }}\left( {{p_1},{p_2},...,{p_{\tilde m}}} \right)} \right\|_2^2$. Then, \eqref{eq:57} can be further transformed as
\begin{subequations}\label{eq:59}
	\begin{align}
	    &\qquad\qquad\qquad\quad\mathop {\min }\limits_{{p_1},{p_2},...,{p_{\tilde m}}} \mathcal{L}_{{\rm{upper}}},\label{eq:59a}\\
		&{\rm{s.t.}}\left\lbrace {{p_1},{p_2},...,{p_{\tilde m}}} \right\rbrace  \subset {{\tilde L}^{\tilde n}},\sum\limits_{i = 1}^{\tilde m} {\rm{Params}} \left( {{p_i}} \right) \geq \mathcal{P}.\label{eq:59b}
	\end{align}
\end{subequations}

However, solving the surrogate objective \eqref{eq:59a} also remains NP-hard. Note that each term in \eqref{eq:59a} represents the MSE between a pruned or unpruned network and the same network with an additional layer pruned. By leveraging the additive property of network layer, where the output distortion caused by multiple perturbations can be approximated as the sum of the distortions caused by each individual perturbation, this additivity is formulated as \cite{xu2023efficient,zhang2024laptop}
\begin{align}\label{eq:60}
	&\mathbb{E}\left\| {{\boldsymbol{\varepsilon }}\left( {{p_1},...,{p_{i - 1}},{p_i}} \right) - {\boldsymbol{\varepsilon }}\left( {{p_1},...,{p_{i - 1}}} \right)} \right\|_2^2 \nonumber \\
    &\!\!\approx \mathbb{E}\left\| {{\boldsymbol{\varepsilon }}\left( {{p_1},...,{p_{i - 2}},{p_i}} \right) - {\boldsymbol{\varepsilon }}\left( {{p_1},...,{p_{i - 2}}} \right)} \right\|_2^2 \nonumber \\
	&\!\!\approx ... \approx \mathbb{E}\left\| {{\boldsymbol{\varepsilon }}\left( {{p_1},{p_i}} \right) - {\boldsymbol{\varepsilon }}\left( {{p_1}} \right)} \right\|_2^2 \approx \mathbb{E}\left\| {{\boldsymbol{\varepsilon }}\left( {{p_i}} \right) - {{\boldsymbol{\varepsilon }}_{{\rm{tea}}}}} \right\|_2^2.
\end{align}
Leveraging \eqref{eq:60}, the surrogate objective in \eqref{eq:59} can be further approximated as 
\begin{subequations}\label{eq:61}
	\begin{align}
		&\quad\quad\quad\mathop {\min }\limits_{{p_1},{p_2},...,{p_{\tilde m}}} \sum\limits_{i = 1}^{\tilde m}\mathbb{E}\left\|  {{{\boldsymbol{\varepsilon }}_{{\text{tea}}}} - {\boldsymbol{\varepsilon }}\left( {{p_i}} \right)} \right\|_2^2,\label{eq:61a}\\
		&\quad{\rm{s.t.}}\left\lbrace {{p_1},{p_2},...,{p_{\tilde m}}} \right\rbrace  \subset {{\tilde L}^{\tilde n}},\sum\limits_{i = 1}^{\tilde m} {\rm{Params}} \left( {{p_i}} \right) \geq \mathcal{P}.\label{eq:61b}
	\end{align}
\end{subequations}
Based on this approximate objective in \eqref{eq:61}, the term $\mathbb{E}\|  {{{\boldsymbol{\varepsilon }}_{{\text{tea}}}} - {\boldsymbol{\varepsilon }}( {{p_i}})} \|_2^2$ acts as the criterion for pruning. Therefore, we only need to compute $\mathbb{E}\|  {{{\boldsymbol{\varepsilon }}_{{\text{tea}}}} - {\boldsymbol{\varepsilon }}( {{\tilde l_i}} )} \|_2^2$, the output loss between the original network and the network with only the ${\tilde l_i}$-th layer pruned, which has a time complexity of $\mathcal{O}( {\tilde n}) $ per layer ${\tilde l_i} \in {\tilde L^{\tilde n}} $, thereby transforming the problem into a variant of the 0-1 Knapsack problem. For a series of 0-1 Knapsack problems, the classical dynamic programming algorithm can be utilized for solving them. Finally, our designed one-shot layer pruning algorithm for the CGDM is summarized in \textbf{\alref{alg:algtrhpf}}. Then, the total time complexity of \textbf{\alref{alg:algtrhpf}} is $\mathcal{O}({{\tilde nU} \mathord{\left/
		{\vphantom {{\tilde n U} {\bar s}}} \right.
		\kern-\nulldelimiterspace} {\bar s}} + \tilde n \mathcal{C})$ and the storage complexity is $\mathcal{O}(\mathcal{C})$, where $U$ is the number of training samples, $\bar s$ is the number of parallel processing processes, and $\mathcal{C}$ is Knapsack capacity.
	
\subsection{Fine-Tuning LiCGDM with Multi-Objective Distillation}
Typically, the performance of the CGDM may degrade after certain layers are removed from the teacher network. To address this, for the pruned CGDM, referred to as LiCGDM, it is necessary to perform further weight readjustment to restore its performance. Therefore, we enhance the reconstruction performance of the LiCGDM by introducing a re-weighting strategy based on knowledge distillation. Specifically, the overall retraining of the LiCGDM is achieved by combining one task objective and two knowledge distillation objectives:
\begin{align}
  {\mathcal{L}_{\rm{KD} }} = {\mathcal{L}_{{\text{Task}}}} + {\lambda _{\text{O}}}{\mathcal{L}_{\text{OKD}}} + {\lambda _{\text{F}}}{L_{\text{FKD}}},
\end{align}
where
\begin{align}
	{\mathcal{L}_{{\text{Task}}}}=\mathbb{E}_{{ \boldsymbol{\dot{\mathsf{G}}}},{{\boldsymbol{\mathsf{G}}}_t},{{\boldsymbol{\varepsilon }}_t}}{\left\| {{{\boldsymbol{\varepsilon }}_t} - {{\boldsymbol{\varepsilon }}_{\boldsymbol{S }}}\left( {{ \boldsymbol{\dot{\mathsf{G}}}} ,{{\boldsymbol{\mathsf{G}}}_t},t} \right)} \right\|_2^2},
\end{align}
\begin{align}
	{\mathcal{L}_{\text{OKD}}}=\mathbb{E}_{{ \boldsymbol{\dot{\mathsf{G}}}},{{\boldsymbol{\mathsf{G}}}_t},t}{\left\| {{{\boldsymbol{\varepsilon }}_{\boldsymbol{T }}}\left( {{ \boldsymbol{\dot{\mathsf{G}}}} ,{{\boldsymbol{\mathsf{G}}}_t},t} \right)- {{\boldsymbol{\varepsilon }}_{\boldsymbol{S }}}\left( {{ \boldsymbol{\dot{\mathsf{G}}}} ,{{\boldsymbol{\mathsf{G}}}_t},t} \right)} \right\|_2^2},
\end{align}
\begin{align}
	{\mathcal{L}_{\text{FKD}}}\!=\!\sum\nolimits_{\text{i}}\mathbb{E}_{{ \boldsymbol{\dot{\mathsf{G}}}},{{\boldsymbol{\mathsf{G}}}_t},t}{\left\| {{{{\mathcal{F}}}^i_{\boldsymbol{T }}}\left( {{ \boldsymbol{\dot{\mathsf{G}}}} ,{{\boldsymbol{\mathsf{G}}}_t},t} \right)- {{{\mathcal{F}}}^i_{\boldsymbol{S }}}\left( {{ \boldsymbol{\dot{\mathsf{G}}}} ,{{\boldsymbol{\mathsf{G}}}_t},t} \right)} \right\|_2^2}.
\end{align}
Here ${{\boldsymbol{\varepsilon }}_{\boldsymbol{T }}}( {{ \boldsymbol{\dot{\mathsf{G}}}} ,{{\boldsymbol{\mathsf{G}}}_t},t})$ and ${{\boldsymbol{\varepsilon }}_{\boldsymbol{S }}}( {{ \boldsymbol{\dot{\mathsf{G}}}} ,{{\boldsymbol{\mathsf{G}}}_t},t})$ represent the outputs of the teacher model (the frozen CGDM) and the student model (the LiCGDM), respectively. ${{{\mathcal{F}}}^i_{\boldsymbol{T }}}( {{ \boldsymbol{\dot{\mathsf{G}}}} ,{{\boldsymbol{\mathsf{G}}}_t},t})$ and ${{{\mathcal{F}}}^i_{\boldsymbol{T }}}( {{ \boldsymbol{\dot{\mathsf{G}}}} ,{{\boldsymbol{\mathsf{G}}}_t},t} )$ denote the feature maps at the end of the $i$-th stage for the CGDM and LiCGDM, respectively. Without any hyperparameter tuning, we set the values of ${\lambda _{\text{O}}}$ and ${\lambda _{\text{F}}}$ to 1.



\section{Numerical Experiment}
In this section, the implementation details are first introduced. Then, we analyze the convergence and complexity of the proposed model under different hyperparameter settings. Finally, we comprehensively evaluate the proposed approach in terms of reconstruction accuracy and knowledge transfer capability, and compare it against several state-of-the-art methods.

\subsection{Implementation Details}
\begin{figure}[!t]
	\centering
	\includegraphics[width=0.396\textwidth]{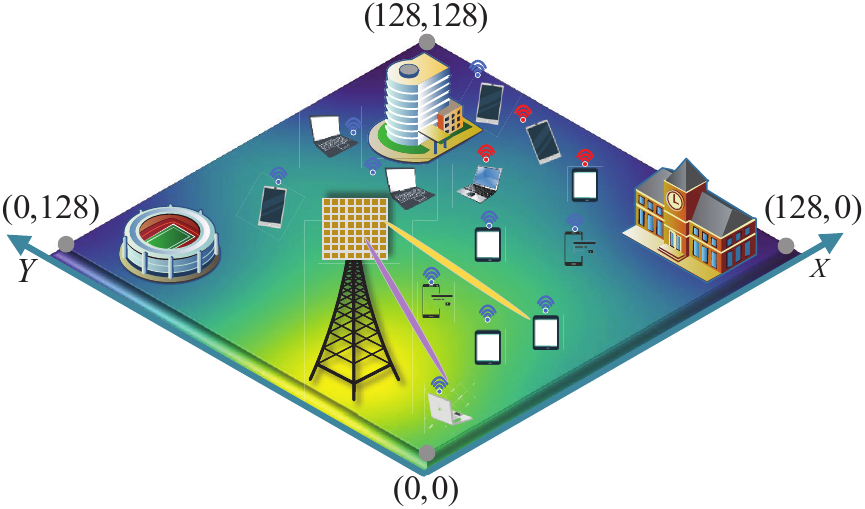}
	\captionsetup{font=footnotesize}
	\caption{The layout of the massive MIMO-OFDM system.}
	\label{fig:layout}
\end{figure}

\textit{i) Wireless Communication Scenario Setup:} The layout of the communication scenario is illustrated in \figref{fig:layout}. Specifically, we consider a massive MIMO-OFDM system operating within a communication area $A$ measuring 128 m $\times$ 128 m, where the BS is equipped with an $8\times8$ UPA with half-wavelength spacing. The ground-truth channels are generated using the widely adopted QuaDRiGa generator (version 2.6.1) \cite{jaeckel2014quadriga}, which employs a geometry-based stochastic channel modeling approach to simulate realistic radio channel impulse responses for mobile radio networks. Meanwhile, we consider the 5G NR typical urban micro-cell scenario ``3GPP 38.901 UMa NLOS", which encompasses both LOS and NLOS physical propagation environments. Additionally, the relevant simulation parameters are summarized in \tabref{ta:sys}.

\textit{ii) CF Data Generation:} In the aforementioned wireless communication scenario, we set ${\sigma}_{\rm{HR}}=128$ and utilize a sampling interval of $\Delta_x=\Delta_y=1$ m to sample the target area $A$ along with its corresponding channel power values, resulting in the HR CF, denoted as $\boldsymbol{\mathsf{G}}_{{\rm{HR}},u}$. For the highly challenging $\times4$ SR CF reconstruction task, ${\sigma}_{\rm{LR}}$ is set to 32, meaning the sampling interval for the LR CF is $\Delta_x = \Delta_y = 4$ m, thereby obtaining the LR CF defined as $\boldsymbol{\mathsf{G}}_{{\rm{LR}},u}$. Similarly, we generate 6,000 pairs of CF samples, denoted as $\{ {\boldsymbol{\mathsf{G}}_{{\rm{HR}},u},\boldsymbol{\mathsf{G}}_{{\rm{LR}},u}} \}_{u = 1}^{6000}$, by running QuaDRiGa simulations with different BS locations $(x,y)$, where $x$ and $y$ are randomly chosen integers in the range $[1,128]$. This data generation method not only preserves the diversity of the dataset, capturing the complexity and dynamic variability inherent in wireless communication environments, but also further validates the generalization performance of the proposed model. These raw samples are subsequently divided into training and testing sets in a 5:1 ratio. To enhance the efficiency of the training process, we apply min-max normalization to the raw CF, i.e., 
\begin{align}\label{eq:}
	{{{{G}}}^{\prime }}\left(\mathfrak{e}, {{\mathbf{\Lambda }}_{i,j}}\right)=\frac{{{G}}\left(\mathfrak{e}, {{\mathbf{\Lambda }}_{i,j}}\right)-\min \left( {{G}}\left(\mathfrak{e}, {{\mathbf{\Lambda }}_{i,j}}\right)\right)}{\max \left( {{G}}\left(\mathfrak{e}, {{\mathbf{\Lambda }}_{i,j}}\right) \right)-\min \left( {{G}}\left(\mathfrak{e}, {{\mathbf{\Lambda }}_{i,j}}\right) \right)}.
\end{align}

\newcolumntype{L}{>{\hspace*{-\tabcolsep}}l}
\newcolumntype{R}{c<{\hspace*{-\tabcolsep}}}
\definecolor{lightblue}{rgb}{0.93,0.95,1.0}
\begin{table}[!b]
	\captionsetup{font=footnotesize}
	\caption{Wireless System and Model Setup Parameters}\label{ta:sys}
	\centering
	\setlength{\tabcolsep}{13mm}
	\ra{1.8}
	\scriptsize
	\scalebox{0.8}{\begin{tabular}{LR}
			\toprule
			Parameter &  Value\\
			\midrule
			\rowcolor{lightblue}
			Size of the interested area $A$ & 128 m$\times$128 m\\
			Number of BS antennas & $N_{r,v}\times N_{r,h}=8 \times 8$  \\
			\rowcolor{lightblue}
			Center frequency & 2.4 GHz\\
			Subcarrier spacing & $\Delta_{f}=15$ kHz\\
			\rowcolor{lightblue}
			Subcarrier & ${N_c}=512$\\
			Active subcarrier & $N_k=300$\\
			\rowcolor{lightblue}
			UE velocity & $0.3$ m/s \\
			BS height & 25 m\\
			\rowcolor{lightblue}
			UE height & 1.5 m\\
			Number of base channels & $c_{{1}}=64 $\\
			\rowcolor{lightblue}
			Numbers of integrated Res$^+$ and self-attention blocks & $N_{{{\rm{RA}}}}=2 $\\
			Channel number multiplier  & $\bar \eta  = 1:2:4:8:16$\\
			\rowcolor{lightblue}
			EMA rate& ${{e_r}}=0.9999$ \\
			Learning rate& ${\rm{lr}}=5 \times {10^{ - 5}}$ \\		
			\rowcolor{lightblue}	
			Batch size &16		\\	
			\bottomrule
		\end{tabular}
	}
\end{table}

\textit{iii) Training Strategy and Model Configuration:} At the hardware level, the proposed CGDM is trained utilizing 2 Nvidia RTX-4090 GPUs, each with 24 GB of memory, and tested on a single Nvidia RTX-4090 GPU with 24 GB of memory. At the algorithm level, we employ the Adam optimizer with a learning rate of $5 \times {10^{ - 5}}$ for model parameter updates over 500,000 iterations, and the batch size is set to 16. Starting from the 5,000th iteration, we introduce the exponential moving average algorithm \cite{Ho11}, with the decay factor set to 0.9999. Additionally, we incorporate dropout, with the dropout rate configured at 0.1. The forward diffusion steps $T$ are set to 1000 and the diffusion noise level adheres to a linear variance schedule, ranging from $\beta _1=10^{-6}$ to $\beta _T= 10^{-2}$. To ensure the model's generalization capability, we forgo checkpoint selection on the CGDM and utilize only the most recent checkpoint. More detailed hyperparameter settings for the model are presented in \tabref{ta:sys}.

\textit{iv) Performance Evaluation Metrics:} For a fair comparison, we employ four widely adopted metrics, namely normalized mean squared error (NMSE), mean squared error (MSE), peak signal-to-noise ratio (PSNR), and structural similarity (SSIM). These metrics are defined as follows: $	{\text{NMSE}} = \frac{{\sum\nolimits_{i = 0}^{{N_x}} {\sum\nolimits_{j = 0}^{{N_y}} {\| {{{\hat G}}(\mathfrak{e}, {{\mathbf{\Lambda }}_{i,j}}) - {{G^{\prime }}}(\mathfrak{e}, {{\mathbf{\Lambda }}_{i,j}})} \|} _2^2} }}{{\sum\nolimits_{i = 0}^{{N_x}} {\sum\nolimits_{j = 0}^{{N_y}} {\| {{{ G^{\prime }}}(\mathfrak{e}, {{\mathbf{\Lambda }}_{i,j}})} \|}_2^2} }}$, ${\text{MSE}} =\frac{1}{{{N}_{x}}\times {{N}_{y}}}\sum\nolimits_{i=0}^{{{N}_{x}}}{{{\sum\nolimits_{j=0}^{{{N}_{y}}}{\| {{{\hat G}}(\mathfrak{e}, {{\mathbf{\Lambda }}_{i,j}}) - {{G^{\prime }}}(\mathfrak{e}, {{\mathbf{\Lambda }}_{i,j}})} \|}_2^2}}}$, $	{\rm PSNR} = 20{{\rm{log}}_{10}}( {\frac{{{255} }}{{\sqrt {{\text{MSE}}} }}} )$, ${\rm SSIM}  = \frac{{( 2{u_{{\hat {\boldsymbol{\mathsf{G}}}}}}{u_{{\boldsymbol{\mathsf{G}}}^{\prime }}} + {C_1}) ( 2{\delta _{{{\hat {\boldsymbol{\mathsf{G}}}}{\boldsymbol{\mathsf{G}}}^{\prime }}}} + {C_2}) }}{{( u_{{\hat {\boldsymbol{\mathsf{G}}}}}^2u_{{\boldsymbol{\mathsf{G}}}^{\prime }}^2 + {C_1}) ( \delta _{{\hat {\boldsymbol{\mathsf{G}}}}}^2\delta _{{\boldsymbol{\mathsf{G}}}^{\prime }}^2 + {C_2}) }}$, where ${N_x} = {N_y} = \sigma$, ${{\hat G}}(\mathfrak{e}, {{\mathbf{\Lambda }}_{i,j}})$ and ${{G^{\prime }}}(\mathfrak{e}, {{\mathbf{\Lambda }}_{i,j}})$ represent the predicted channel channel power and the ground-truth channel power, respectively. ${\boldsymbol{\mathsf{G}}}^{\prime }$ is the input LR CF, ${\hat {\boldsymbol{\mathsf{G}}}}$ is the reconstructed HR CF, ${{u_{{\hat {\boldsymbol{\mathsf{G}}}}}}}$ and ${{u_{{\boldsymbol{\mathsf{G}}}^{\prime }}}}$ are the means, ${\delta _{{\hat {\boldsymbol{\mathsf{G}}}}}^2}$ and ${\delta _{{\boldsymbol{\mathsf{G}}}^{\prime }}^2}$ are the variances of ${{\hat {\boldsymbol{\mathsf{G}}}}}$ and ${{\boldsymbol{\mathsf{G}}}^{\prime }}$, respectively. ${{\delta _{{{\hat {\boldsymbol{\mathsf{G}}}}{\boldsymbol{\mathsf{G}}}^{\prime }}}}}$ is covariance of ${{\hat {\boldsymbol{\mathsf{G}}}}}$ and ${{\boldsymbol{\mathsf{G}}}^{\prime }}$, $C_1$ and $C_2$ represent nonzero constants.

\subsection{Experiment Results} \label{Experiment Results}
\textit{i) Convergence and Complexity Analysis:}
According to the proposed CGDM architecture shown in \figref{fig:model}, the performance of the CGDM relies on the number $c_1$ of base channels in the feature maps within the latent space, as well as the number $N_{\rm{RA}}$ of integrated modules that combine the Res$^+$ block and the self-attention block. Additionally, during the model training process, the depth of the feature maps, namely the number of channels, also affects the CGDM's ability to represent features. The channel number multiplier is defined as $\bar \eta  = {c_1}:{c_2}:...:{c_{\bar n}}$, where ${\bar n}$ is related to the number of down-sampling operations. Consequently, we analyze the convergence and complexity of CGDM under different influencing factors. Note that the training data used for the analysis in this subsection are all derived from the $32\times32 \to  128\times128$ HR CF reconstruction task. The default parameter configuration for the CGDM is $c_1=64$, $\bar \eta  = 1:2:4:8:16$, and $N_{\rm{RA}}=2$. Any parameters not explicitly mentioned in the following analysis are assumed to be set to these values.

\begin{figure*}[!t]
	\centering
	\subfloat[]{\includegraphics[width=0.33\textwidth]{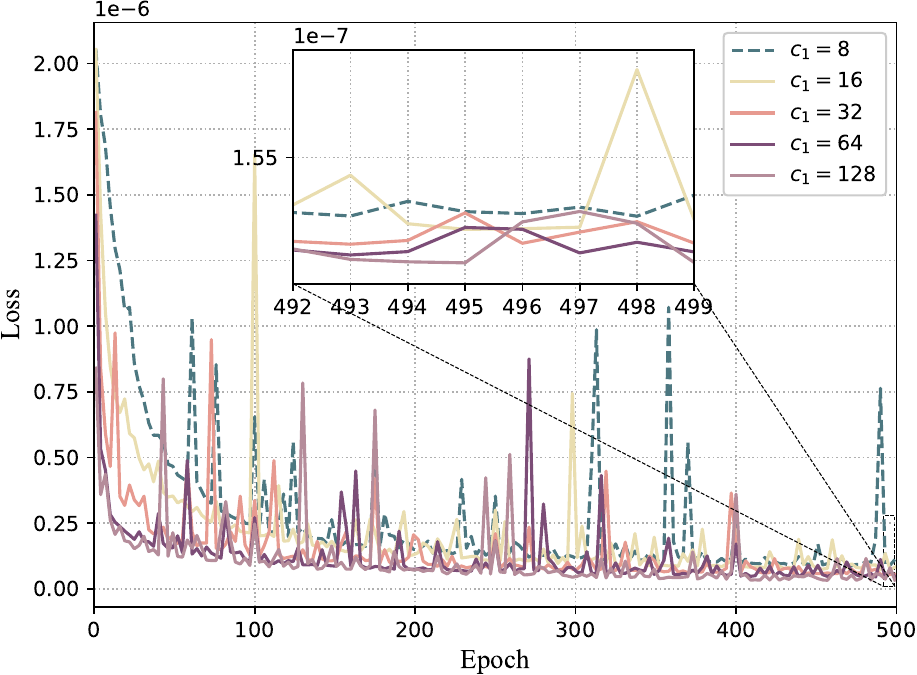}\label{fig:c1}}
	\hfill
	\subfloat[]{\includegraphics[width=0.33\textwidth]{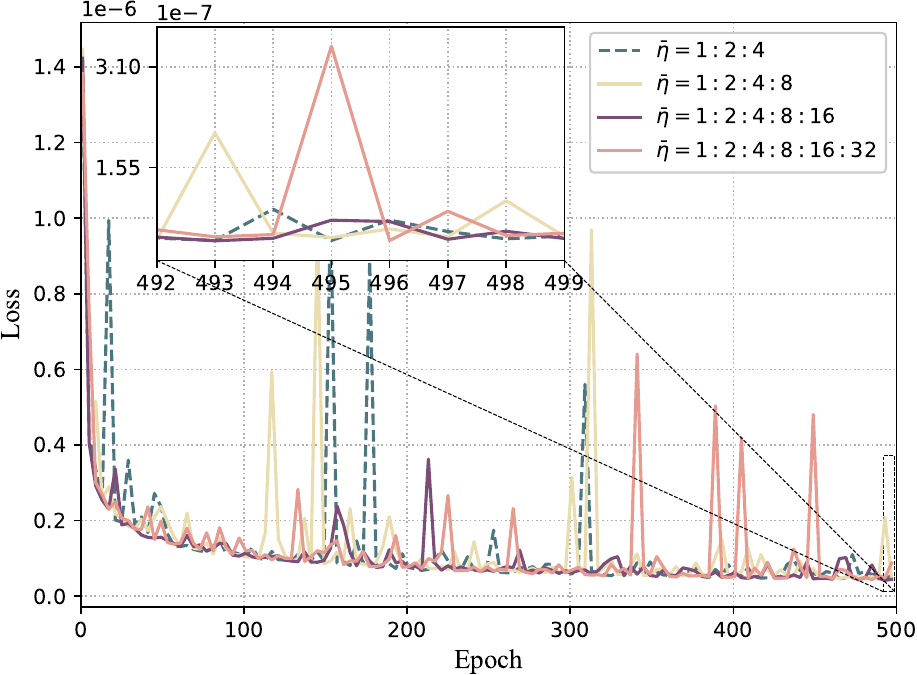}\label{fig:eta}}
	\hfill
	\subfloat[]{\includegraphics[width=0.33\textwidth]{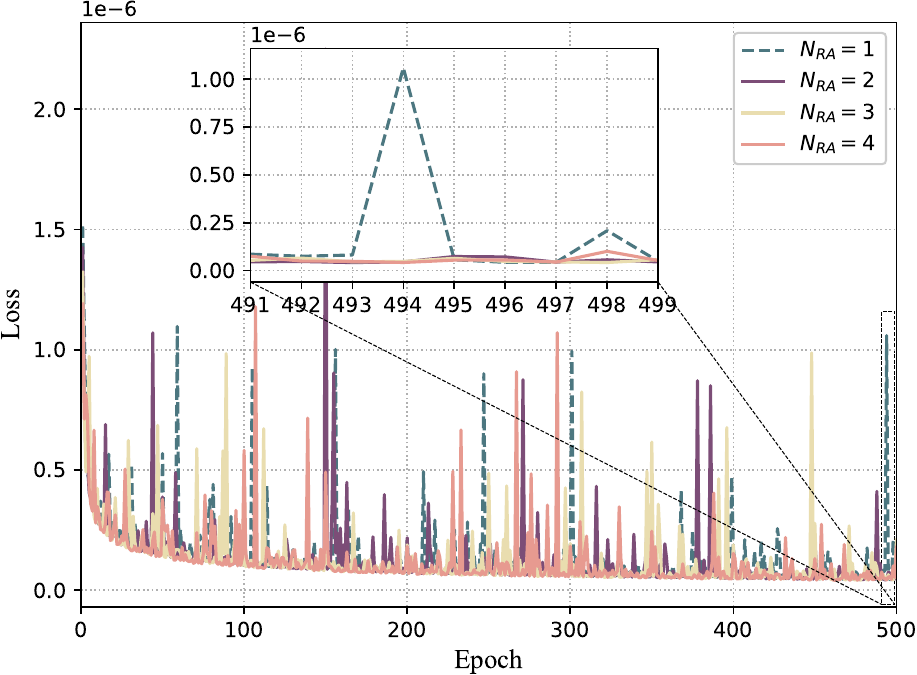}\label{fig:NRA}}
	\captionsetup{font=footnotesize}
	\caption{Convergence analysis of the CGDM under different hyperparameter settings: (a) represents different base channels $c_1$ in the feature maps within the latent space; (b) shows different channel number multipliers $\bar \eta$; (c) illustrates varying numbers $N_{\rm{RA}}$ of integrated Res$^+$ and self-attention blocks.}
	\label{fig:loss}
\end{figure*}

As shown in \subfigref{fig:loss}{fig:c1}, the variation of the loss function for the CGDM with different numbers, $c_1$, of base channels, is presented over 500 training epochs. It can be observed that the CGDM loss corresponding to $c_1=\left\lbrace 8,16,32,64,128\right\rbrace $ decreases to a relatively low value, specifically on the order of $10^{-6}$, within the first 100 epochs. For better clarity, we further zoom in on the loss between epochs 490 and 500. It can be observed that the loss corresponding to $c_1=\left\lbrace 8,16,32,64,128\right\rbrace$ for CGDM fluctuates at the order of $10^{-7}$, indicating a very slight loss oscillation. Additionally, an interesting observation is that by appropriately increasing $c_1$, the overall amplitude of the loss oscillations during the CGDM training process decreases, while the loss itself also reduces. Therefore, setting $c_1=64$ is a suitable choice for our task.
\begin{figure}[!b]
	\centering
	\subfloat[]{\includegraphics[width=0.46\textwidth]{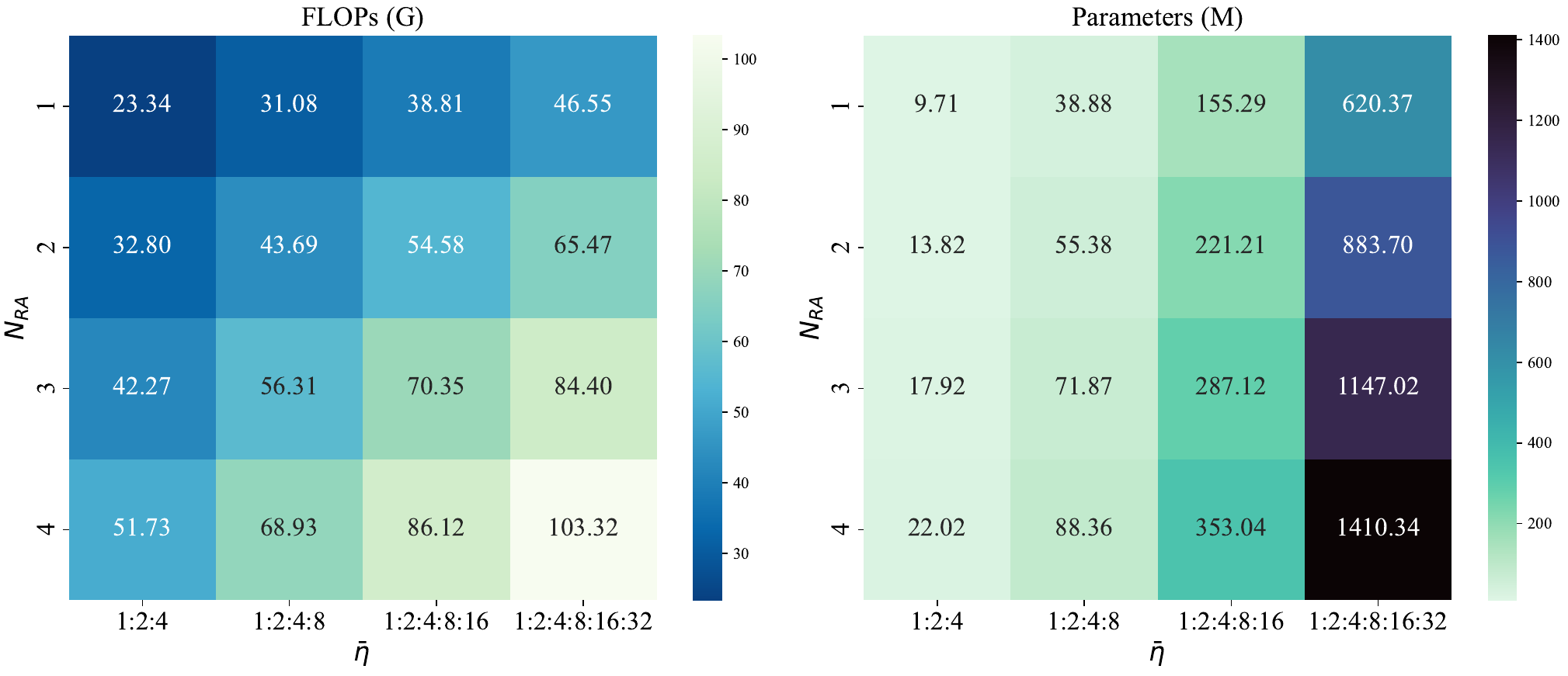}\label{fig:c1=64}}\\
	\subfloat[]{\includegraphics[width=0.46\textwidth]{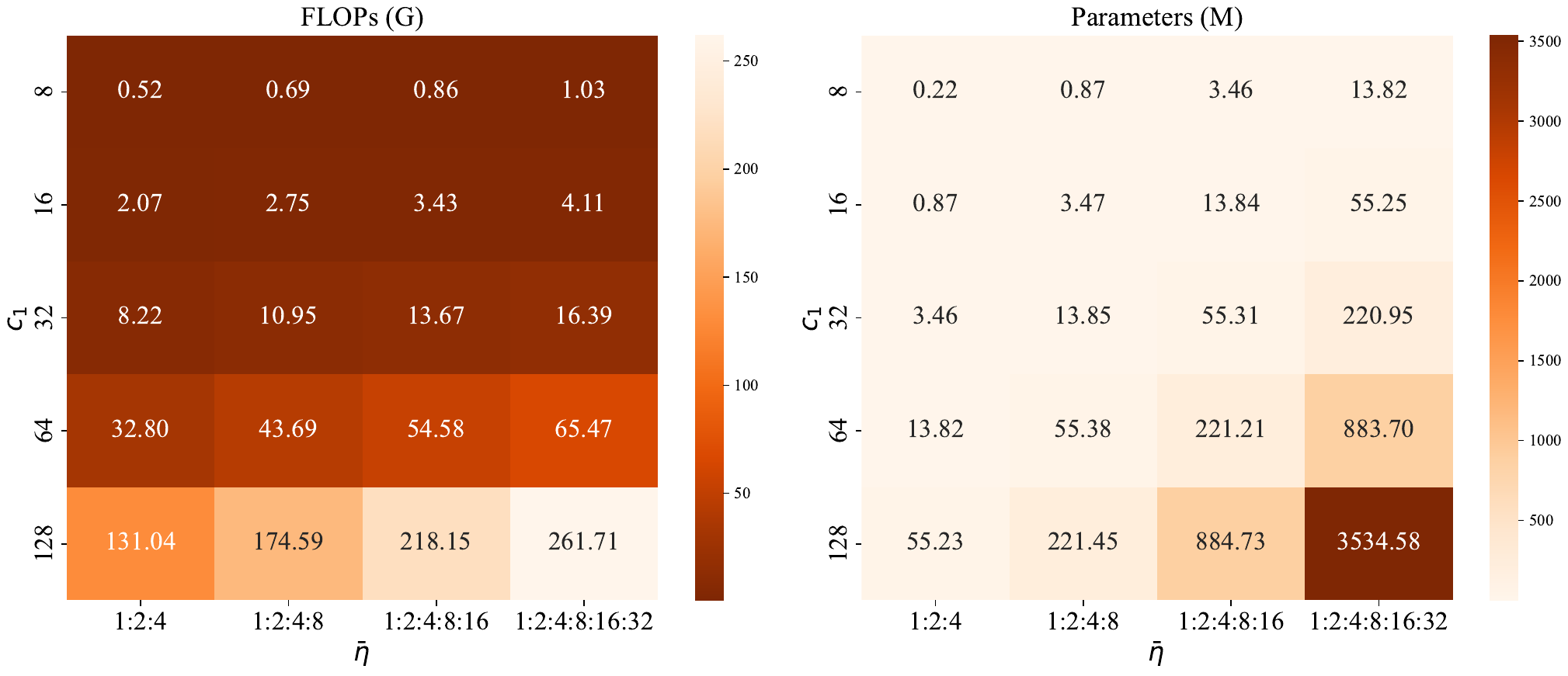}\label{fig:nra=2}}
	\captionsetup{font=footnotesize}
	\caption{Analysis of the computational complexity (FLOPs in Giga) and storage complexity (Parameters in Millions) of CGDM under different parameter settings. (a) shows the impact of different settings for $N_{\rm{RA}}$ and $\bar \eta$ on the CGDM complexity when $c_1$ is set to 64. (b) illustrates the impact of different settings for $c_1$ and $\bar \eta$ on the CGDM complexity when $N_{\rm{RA}}$ is set to 2.}
	\label{fig:flops}
\end{figure}

As shown in \subfigref{fig:loss}{fig:eta}, the variation of the loss function for CGDM with different channel number multiplier $\bar \eta$ over 500 training epochs is presented. The CGDM, with different $\bar \eta$ configurations, converges to a minimal value on the order of $10^{-6}$ within the first 100 epochs. Additionally, between epochs 490 and 500, the variation remains within a narrow range, on the order of $10^{-7}$, indicating minimal fluctuation. However, a shallower channel number multiplier, such as $\bar \eta=1:2:4$, tends to cause relatively larger fluctuations in the loss value. While a deeper channel number multiplier can progressively extract more high-level features, it also introduces challenges in training and optimization, as seen with $\bar \eta=1:2:4:8:16:32$. Therefore, setting $\bar \eta=1:2:4:8:16$ is a reasonable choice for our task. \subfigref{fig:loss}{fig:NRA} shows the variation of the CGDM loss function as the number of integrated Res$^+$ and self-attention modules increases. Overall, for CGDM with $N_{\rm{RA}}$ set to $\left\lbrace 1,2,3,4\right\rbrace $, the loss value decreases to a small value within the first 500 epochs, with minimal fluctuation, and tends to converge. To maintain the model's ability to represent features and ensure effective interaction between local and global features, we choose $N_{\rm{RA}}=2$ as a suitable option.

The complexity of a large AI model is primarily determined by two key factors: the number of model parameters and the number of floating point operations (FLOPs). For clarity, we visualize the impact of varying hyperparameters $N_{\rm{RA}}$, $c_1$, and $\bar \eta$ on the complexity of the CGDM. Specifically, \subfigref{fig:flops}{fig:c1=64} illustrates the variation in CGDM complexity under different settings of the number $N_{\rm{RA}}$ of integrated Res$^+$ and attention modules, as well as channel number multipliers $\bar \eta$, with $c_1$ fixed at 64. \subfigref{fig:flops}{fig:nra=2} shows the variation in CGDM complexity under different settings of the base channel number $c_1$ and channel number multiplier $\bar \eta$, with $N_{\rm{RA}}$ fixed at 2. Overall, appropriately increasing these parameters facilitates the extraction of higher-level features and multi-scale fusion in CGDM, while also raising the hardware requirements for model deployment. Therefore, there is a trade-off between CGDM reconstruction performance and complexity. Based on the convergence and complexity analysis, we set $c_1=64$, $N_{\rm{RA}}=2$, and $\bar \eta=1:2:4:8:16$ as the default configuration for subsequent experiments, with the corresponding parameter count of 221.21 M and FLOPs of 54.58 G.







\textit{ii) Quantitative and Qualitative Comparison with Baselines:} To ensure a fair comparison, we apply the same training strategy as CGDM to the baselines, conducting weight learning and testing on the same $\times4$ CF dataset. To comprehensively evaluate the performance of the proposed CGDM and LiCGDM (with $35\% $ pruning), we compare them with several state-of-the-art models, including SRGAN-MSE \cite{ledig2017photo}, SRGAN \cite{ledig2017photo}, C-VAE \cite{kingma2013auto}, and DRRN \cite{tai2017image}. 
\tabref{Performance comparison} provides a quantitative performance analysis of the proposed model compared to baseline models on $\times4$ HR CF reconstruction tasks. The proposed CGDM demonstrates competitive performance across four metrics (NMSE, MSE, PSNR, and SSIM) and outperforms the baselines. Specifically, compared to SRGAN, CGDM reduces NMSE and MSE $65.236\times 10^{-5}$ and 9.542, respectively, while improving PSNR and SSIM by 11.747 and 0.006, respectively. LiCGDM is a student model derived from CGDM, achieved by pruning $35\%$ of the parameters and fine-tuning the remaining ones. Compared to CGDM, LiCGDM exhibits a slight decrease in performance metrics, but the degradation remains within an acceptable range. This confirms that the proposed lightweight approach effectively compresses CGDM, enabling its practical deployment on personal computers and even mobile devices.

\begin{figure*}[!t] 
	\centering
	\includegraphics[width=0.966\textwidth]{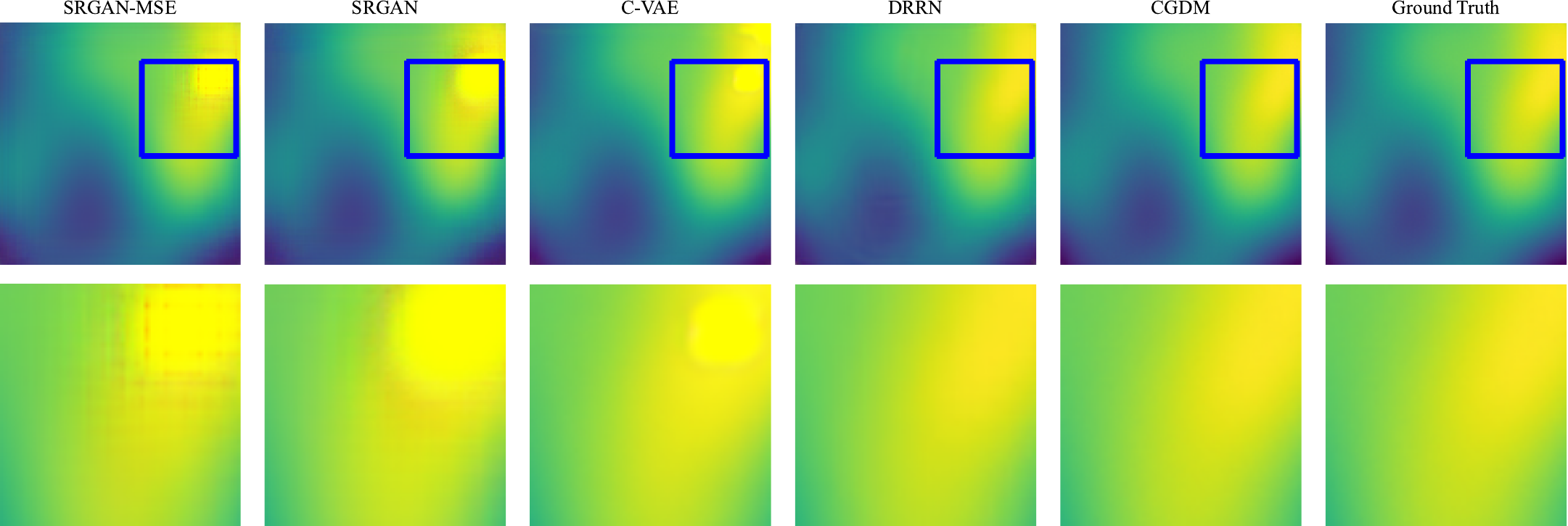}
	\captionsetup{font=footnotesize}
	\caption{Qualitative comparison in the $\times4$ HR CF reconstruction task. The first row displays the reconstruction results of the proposed CGDM and baseline models. For clarity, the second row shows zoomed-in views of the regions highlighted by the blue boxes.}
	\label{fig:srdiff}
\end{figure*}

For the qualitative analysis, we visualize the results of the proposed CGDM and baselines on the $32 \times 32 \to 128 \times 128$ HR CF reconstruction task, as shown in \figref{fig:srdiff}. As observed, both our model and DRRN produce clearer and more accurate pattern edges, while other baselines tend to generate blurrier results that deviate from the ground-truth target.

\newcolumntype{L}{>{\hspace*{-\tabcolsep}}l}
\newcolumntype{R}{c<{\hspace*{-\tabcolsep}}}
\definecolor{lightblue}{rgb}{0.93,0.95,1.0}
\begin{table}[!t]
	\captionsetup{font=footnotesize}
	\caption{Quantitative Evaluation of the Proposed Approach and Baselines on NMSE, MSE, PSNR, and SSIM for the $\times4$ SR CF.}\label{Performance comparison}
	\centering
	\setlength{\tabcolsep}{1.7mm}
	\ra{1.8}
	\scriptsize
	\scalebox{0.9}{\begin{tabular}{cccccR}
			\toprule
			\multicolumn{1}{c}{Reconstruction Task}                & \multicolumn{1}{c}{Method}     & \multicolumn{1}{c}{ NMSE ($\times 10^{-5}$)}           & \multicolumn{1}{c}{MSE}             & \multicolumn{1}{c}{PSNR}       & \multicolumn{1}{R}{SSIM}      \\ \midrule
			{\multirow{5}{*}{\begin{tabular}[c]{@{}c@{}}$\times4$\\ $32^2\to 128^2$\end{tabular}}}  & 	SRGAN-MSE \cite{ledig2017photo}       & 62.049      & 9.060     & 38.610       & 0.987  \\
			& \cellcolor{lightblue}SRGAN \cite{ledig2017photo}         & \cellcolor{lightblue}69.942              & \cellcolor{lightblue}10.220              & \cellcolor{lightblue}38.104        & \cellcolor{lightblue}0.987          \\
			& C-VAE \cite{kingma2013auto}          & 59.833     & 8.701          & 39.022     & 0.989    \\
			& \cellcolor{lightblue}DRRN \cite{tai2017image}          & \cellcolor{lightblue}65.978     & \cellcolor{lightblue}8.656         & \cellcolor{lightblue}41.009    & \cellcolor{lightblue}0.988     \\
			&CGDM (Our)        & \textbf{4.706}$\downarrow$          & \textbf{0.678}$\downarrow$           & \textbf{49.851}$\uparrow$        & \textbf{0.993}$\uparrow$           \\
			& \cellcolor{lightblue}LiCGDM (Our)        & \cellcolor{lightblue}6.653          & \cellcolor{lightblue}0.9640          & \cellcolor{lightblue}49.726      & \cellcolor{lightblue}0.993 
			  \\  	\bottomrule
		\end{tabular}
	}
\end{table}
\begin{table}[!t]
	\captionsetup{font=footnotesize}
	\caption{Zero-Shot Performance Comparison of the Proposed Approach and Baselines on $\times16$, $\times8$, and $\times2$ HR CF Reconstruction Tasks}\label{zero-shot}
	\centering
	\setlength{\tabcolsep}{1.7mm}
	\ra{1.8}
	\scriptsize
	\scalebox{0.9}{\begin{tabular}{cccccR}
			\toprule
			\multicolumn{1}{c}{Zero Shot}                & \multicolumn{1}{c}{Method}     & \multicolumn{1}{c}{ NMSE ($\times 10^{-5}$)}           & \multicolumn{1}{c}{MSE}             & \multicolumn{1}{c}{PSNR}       & \multicolumn{1}{R}{SSIM}      \\ \midrule
			{\multirow{5}{*}{\begin{tabular}[c]{@{}c@{}}$\times16$\\ $8^2\to 128^2$\end{tabular}}}  & 	SRGAN-MSE \cite{ledig2017photo}       & 3031.40     & 437.077     & 21.812      & 0.741  \\
			& \cellcolor{lightblue}SRGAN \cite{ledig2017photo}         & \cellcolor{lightblue}3440.20             & \cellcolor{lightblue}497.045            & \cellcolor{lightblue}21.270        & \cellcolor{lightblue}0.728         \\
			& C-VAE \cite{kingma2013auto}          & 660.69    & 96.170          & 28.505     & \textbf{0.860}$\uparrow$   \\
			& \cellcolor{lightblue}DRRN \cite{tai2017image}          & \cellcolor{lightblue}1127.20     & \cellcolor{lightblue}157.350         & \cellcolor{lightblue}26.612    & \cellcolor{lightblue}0.844     \\
			&CGDM (Our)         & \textbf{588.82}$\downarrow$          & \textbf{85.667}$\downarrow$           & \textbf{29.023}$\uparrow$        & 0.855           \\
			& \cellcolor{lightblue}LiCGDM (Our)       & \cellcolor{lightblue}601.78         & \cellcolor{lightblue}87.572         & \cellcolor{lightblue}28.192      & \cellcolor{lightblue}0.855 
			\\ \midrule
			{\multirow{5}{*}{\begin{tabular}[c]{@{}c@{}}$\times8$\\ $16^2\to 128^2$\end{tabular}}}  & 	SRGAN-MSE \cite{ledig2017photo}       & 560.88     & 80.357     & 29.132       & 0.867  \\
			& \cellcolor{lightblue}SRGAN \cite{ledig2017photo}         & \cellcolor{lightblue}695.55             & \cellcolor{lightblue}99.956             & \cellcolor{lightblue}28.220        & \cellcolor{lightblue}0.858         \\
			& C-VAE \cite{kingma2013auto}          & 103.68     & 15.038          & 36.492     & \textbf{0.986}$\uparrow$     \\
			& \cellcolor{lightblue}DRRN \cite{tai2017image}          & \cellcolor{lightblue}265.94     & \cellcolor{lightblue}36.083        & \cellcolor{lightblue}33.764    & \cellcolor{lightblue}0.940   \\
			&CGDM (Our)         & \textbf{67.84}$\downarrow$          & \textbf{9.866}$\downarrow$           & \textbf{38.393}$\uparrow$        & 0.957          \\
			& \cellcolor{lightblue}LiCGDM (Our)       & \cellcolor{lightblue}78.36         & \cellcolor{lightblue}11.413          & \cellcolor{lightblue}37.719      & \cellcolor{lightblue}0.957 
			\\ 	\midrule
			{\multirow{5}{*}{\begin{tabular}[c]{@{}c@{}}$\times2$\\ $64^2\to 128^2$\end{tabular}}}  & 	SRGAN-MSE \cite{ledig2017photo}       & 1381.70     & 203.691     & 25.544       & 0.917  \\
			& \cellcolor{lightblue}SRGAN \cite{ledig2017photo}         & \cellcolor{lightblue}1343.10            & \cellcolor{lightblue}199.048            & \cellcolor{lightblue}25.709       & \cellcolor{lightblue}0.914      \\
			& C-VAE \cite{kingma2013auto}          & 62.50     & 9.081        & 38.852    & 0.989   \\
			& \cellcolor{lightblue}DRRN \cite{tai2017image}          & \cellcolor{lightblue}66.83    & \cellcolor{lightblue}8.727         & \cellcolor{lightblue}41.271    & \cellcolor{lightblue}0.990  \\
			&CGDM (Our)         & \textbf{15.55}$\downarrow$          & \textbf{2.259}$\downarrow$           & \textbf{44.787}$\uparrow$        & \textbf{0.994}$\uparrow$           \\
			& \cellcolor{lightblue}LiCGDM (Our)       & \cellcolor{lightblue}21.36          & \cellcolor{lightblue}3.113          & \cellcolor{lightblue}44.415     & \cellcolor{lightblue}0.994 
			\\ 	\bottomrule
		\end{tabular}
	}
\end{table}

\textit{iii) Evaluation of Knowledge Transfer and Generalization Ability:} Transferring a trained model to an unseen task is considered zero-shot generation. Namely, the neural network learns and updates its weights solely on the $\times4$ SR CF dataset, without exposure to the $\times16$, $\times8$, and $\times2$ SR CF datasets. It relies on the trained model to perform SR CF reconstruction tasks for magnification factors that have not been encountered before. This is an effective way to assess the model's knowledge transferability and generalization ability. \tabref{zero-shot} summarizes the zero-shot results of the proposed model and baselines on the $\times16$, $\times8$, and $\times2$ reconstruction tasks. It can be observed that across the three unseen SR CF reconstruction tasks, SRGAN and SRGAN-MSE perform slightly worse, while C-VAE achieves slightly better results. We think this is due to the fact that GAN-based models need to iteratively minimize the generator loss and maximize the discriminator loss, which may lead to mode collapse and prevent the models from fully capturing the diversity of the true distribution. In contrast, VAE explicitly estimates the latent parameters by maximizing a lower bound on the log-likelihood. Its mathematical formulation ensures a tractable likelihood for evaluation and enables an explicit inference network. It is worth noting that both the proposed CGDM and LiCGDM outperform the baselines across multiple performance metrics, achieving impressive reconstruction results. We attribute this to CGDM's robust ability to learn implicit priors and its capacity to model complicated data distributions. By leveraging source information and implicit priors, CGDM achieves impressive HR CF reconstruction results.

\section{Conclusion}
To facilitate the paradigm shift from environment-unaware to intelligent and environment-aware communication, this paper introduced the concept of CF twins. Particularly, we treated the coarse-grained and fine-grained CFs as physical and virtual twin objects, respectively, and designed a CGDM as the core computational unit of the CF twins to model their connection. The trained CGDM, combining the learned prior distribution of the target data and side information, generates fine-grained CFs through a series of iterative refinement steps. Additionally, to facilitate the practical deployment of the CGDM, we introduced a one-shot pruning approach and employed multi-objective knowledge distillation techniques to minimize performance degradation. Experimental results showed that the proposed model achieved competitive reconstruction accuracy in fine-grained CF reconstruction tasks with magnification factors of $\times 2$, $\times 4$, $\times 8$, and $\times 16$, while also demonstrating exceptional knowledge transfer capabilities.

\bibliographystyle{IEEEtran}
\bibliography{EE_AI}

\end{document}